\documentclass[%
 reprint,
superscriptaddress,
 amsmath,amssymb,
 aps,
,prd,longbibliography]{revtex4-1}

\usepackage{graphicx}
\usepackage{dcolumn}
\usepackage{comment}
\usepackage{bm}
\usepackage{color}

\begin{document}

\title{Self-assembly and non-equilibrium phase coexistence in a binary granular mixture}
\author{A. Plati}
\affiliation{Universit\'e Paris-Saclay, CNRS, Laboratoire de Physique des Solides, 91405 Orsay, France}
\author{R. Maire}
\affiliation{Universit\'e Paris-Saclay, CNRS, Laboratoire de Physique des Solides, 91405 Orsay, France}
\author{ F. Boulogne}
\affiliation{Universit\'e Paris-Saclay, CNRS, Laboratoire de Physique des Solides, 91405 Orsay, France}
\author{ F. Restagno}
\affiliation{Universit\'e Paris-Saclay, CNRS, Laboratoire de Physique des Solides, 91405 Orsay, France}
\author{ F. Smallenburg}
\affiliation{Universit\'e Paris-Saclay, CNRS, Laboratoire de Physique des Solides, 91405 Orsay, France}
\author{G. Foffi}
\affiliation{Universit\'e Paris-Saclay, CNRS, Laboratoire de Physique des Solides, 91405 Orsay, France}

\date{\today}

\begin{abstract}

We report the experimental observation of a square crystalline phase in a vibrated binary mixture of spherical grains. This structure spontaneously forms from a disordered state, consistently with predictions obtained in an equilibrium system with similar geometrical properties under conservative dynamics. By varying the area fraction, we also observe stable coexistence between a granular fluid and an isolated square crystal. Using realistic simulations based on the discrete element method and an idealized collisional model integrated via event-driven molecular dynamics, we not only reproduce experimental results but also help to gain further insights into the non-equilibrium phase coexistence. Through the direct phase coexistence method, we demonstrate that the system shows behavior highly similar to an equilibrium first-order phase transition. However, the crystal remains at a higher granular temperature than the fluid, which is a striking non-equilibrium effect.
Through qualitative arguments
and supported by kinetic theory, we elucidate the role of the coupling between local structure and
energy transfer mechanisms in sustaining kinetic temperature gradients across the fluid-solid interface. 
\end{abstract}

\maketitle


\section{Introduction}\label{sec:ABSrealistic}
The study of spontaneous self-assembly of ordered structures starting from disordered phases unites the interests of a diverse scientific community ranging from physicists and chemists to engineers \cite{Whitesides2002,vanBlaaderen1997,Cai2021,Boles2016,glotzer2007,Grzybowski2017}.  The paradigmatic example of hard-sphere crystallization demonstrates how equilibrium statistical mechanics can provide a general theoretical framework to understand self-assembly processes in realistic systems such as colloidal suspensions \cite{Royall2023}. Following equilibrium thermodynamics, the spontaneous tendency of an amorphous system to form ordered structures can be understood in terms of phase stability: In a certain region of the control parameter space, ordered macrostates are more thermodynamically stable than disordered ones (i.e. they minimize the system's free energy). In many cases, this results in a first-order phase transition which also implies a coexistence region between the two phases.  Remarkably, self-assembly of ordered structures and phase coexistence have been also observed in systems that are far from thermodynamic equilibrium such as active \cite{Cates2015,Mandal2019,Hecht2024,Omar2021,Takatori2015,bialke2015negative,Caprini2019} and granular matter \cite{Olafsen98,Prevost2004,Reyes2008,Lobkovsky2009,Reis2006,Roeller2011,Clewett2016}. These are intrinsically open systems where energy constantly flows in and out. Thus, despite the possibility for them to attain the so-called non-equilibrium stationary states, where the statistical properties of macroscopic observables do not change with time,  any notion of thermodynamic stability based on free energy arguments is lost. Nonetheless, out-of-equilibrium self-assembly processes can be extremely similar to equilibrium ones in geometrically equivalent conditions. A striking example of that is provided by the crystallization of quasi-2D vibrated granular matter \cite{Olafsen2005,Reis2006,Castillo2012, Castillo2015,Plati2024}. It has been known for more than two decades that a monolayer of monodisperse spherical grains vibrated on a substrate can self-assemble into a hexagonal crystal \cite{Olafsen2005,Reis2006}, and recently it was observed that bidisperse mixtures can spontaneously form a quasi-crystalline structure \cite{Plati2024}. In these examples, the conditions for the liquid-solid transition and the evolution of the structure resemble those of equivalent 2D hard-disk systems undergoing conservative dynamics. Nevertheless, dynamical observables show clear deviations from equilibrium phenomenology, such as non-Maxwellian velocity distributions \cite{Reis2007,Plati2024}. Additionally, the behavior of the granular crystallization turns out to be sensitive to variations in the frequency and the amplitude of the vibration, or the material properties of the grains. Relevant examples are: i) the observation of a non-equilibrium phase coexistence between a ``hot" granular fluid and a ``frozen" crystal \cite{Olafsen98,Melby2005}, ii) a change from a two-step liquid-solid phase transition involving an intermediate hexatic phase to a one-step first-order-like scenario \cite{Komatsu2015} and iii) the suppression of ordering induced by strong inelasticity \cite{Reyes2008}.
\\
Based on these observations, it is clear that we cannot simply conclude that granular quasi-2D systems behave in the same way as equilibrium hard-disks, nor can we entirely dismiss the possibility of predicting their behavior from equilibrium arguments. Given the importance of granular matter for many applications \cite{Jaeger96,Knowlton94,Coussot2005,Li2014,Boechler2011,Wu2019,Karuriya2022} and the lack of a general theoretical framework for its description, understanding which physical conditions are suitable to be treated with  effective equilibrium-like theories represents an intriguing open problem.
In this paper, building on our earlier work on granular quasicrystals \cite{Plati2024}, we tackle this problem by studying experimentally and numerically a vibrated granular system able to self-assemble into a periodic binary crystal with square symmetry. Specifically, we demonstrate self-assembly of square crystals and stable fluid-crystal coexistences in both experiments and simulations of binary mixtures of granular beads. In the realm of granular materials, one usually expects polydisperse mixtures to undergo size segregation effects \cite{Aumaitre2001,WILLIAMS1976,AHMAD1973,BRIDGWATER2012,Rivas_2011,Huerta2004} while any crystalline structure built upon more than one particle species imply a homogeneous concentration of grains with different sizes. In light of this, a promising by-product of our study is the application of self-assembly of binary crystals as a way to realize homogeneous granular mixtures \cite{nakai2024} even in the presence of strong non-equilibrium effects that would promote size segregation. To our knowledge, the only previous example of periodic binary granular crystals obtained in experiments relies on the presence of electrostatic interactions between the grains \cite{Grzybowski2003}, while the dynamics of our system are purely collision-driven. 
We will also see that our system is particularly well suited to the study of non-equilibrium phase coexistence, a subject which has attracted considerable interest in recent years thanks to a series of experimental and numerical observations \cite{Mandal2019,Olafsen98,Prevost2004,Roeller2011,Hecht2024}. 
A striking feature of non-equilibrium phase coexistence is the possibility of maintaining stable kinetic temperature gradients between the two different phases. This suggests the presence of a net heat-flux which is not compatible with thermodynamic equilibrium. 
Kinetic temperature differences in gas-solid and liquid-solid coexistence have been observed in granular \cite{Olafsen98,Prevost2004} and active \cite{Mandal2019,Hecht2024} matter. Here, equilibrium-like general principles able to predict their occurrence and describe their characteristics are absent: Each specific case needs to be analyzed to understand the key physical ingredients that make the coexistence possible and determine its properties such as, for example, which of the two phases is hotter. Through a numerical analysis, we show that in our systems the crystalline phase has a higher granular temperature than the coexisting liquid one. We explain this observation by using the relationship between the system's local structure and energy transfer mechanisms. Remarkably, we find that the kinetic theory of binary granular mixtures is able to predict this phenomenon in a simplified granular model. 

The remainder of this paper is structured as follows. In Sec.~\ref{sec: exp}, we discuss our experimental results about the self-assembly of the granular binary crystal and its stable
coexistence with a granular liquid phase. In Sec.~\ref{sec: simu}, we focus on numerical results obtained following two complementary approaches: realistic simulations based on the Discrete Element Methods (DEM) and Event-Driven Molecular Dynamics (EDMD) of a simplified 2D coarse-grained granular model. Using these simulations, we first confirm and extend the study on self-assembly, and then, also with the help of kinetic theory, we perform a systematic study of the granular temperature in the coexistence region.   In Sec.~\ref{sec: conc} we draw our conclusions and outline possible future developments of the present study.

\section{Experiments}\label{sec: exp}

\subsection{Experimental setup and preliminary study}

\begin{figure}
\includegraphics[width=0.99\columnwidth,clip=true]{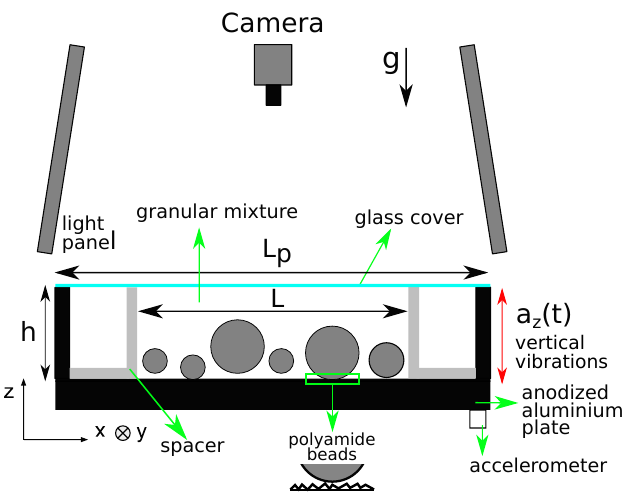}
    \centering
    \caption{Sketch of the experimental setup. A binary mixture of polyamide grains 
    lies on a sandblasted anodized aluminium plate, which is vertically vibrated. Four aluminium spacers are screwed
    on the plate to confine the motion of the grains in a square area of
    side $L = 15$ cm through lateral walls of height $h = 6$ mm while the full lateral side of the plate is $L_p=20$  cm. The granular
    system is confined from above by a glass cover which allows for the direct visualization of the system through a camera. 
    To ease the imaging of the grains, four light panels are placed
    around the apparatus. The measurements of the plate acceleration
    are made possible by a one-axis accelerometer (Brüel \& Kjær, Type
    4534-B-001) that can be rigidly joined to its bottom side. The base of
    the apparatus is held up by three leveling feet  (not shown), which allow for horizontal adjustment.} 
    \label{fig:EXP_setup}
\end{figure}

Our experimental system (see Fig.~\ref{fig:EXP_setup}) consists of a binary mixture of spherical polyamide beads (PROLABO) composed of $N_S$ grains with diameter $\sigma_S=2$ mm and $N_L$ with diameter $\sigma_L=4$ mm. Polyamide has a density $\rho=1.14$ g/cm$^3$. The system is confined in a quasi-2D square box (height $h$ and width $L\gg h$) which is vertically vibrated by an electrodynamic shaker (Brüel \& Kjær, LDS V400). The bottom horizontal surface of the box (1 cm thick) is made of anodized aluminium and, as suggested by a previous study \cite{Reis2007}, has been sandblasted (glass sand, granulometry [425-600] $\mu$m) before anodization to improve grain mobility in the horizontal plane. The upper plate (2 mm thick) is made of glass which allows for the direct visualization of the 2D projection of the system. Snapshots of the horizontal spatial configurations can be acquired with a high-resolution (5 MP) camera (Basler a2A2590) placed at the top of the setup. Unless differently stated, we take a snapshot every 30 seconds. The camera is equipped with a lens (Basler C125-1218-5M-P) having a fixed focal length of 12.0 mm. A signal generator (Keysight 33220A) is used to design a desired waveform $a_z(t)$ which fixes the vertical acceleration of the box
(more details on the experimental setup and procedure are given in the caption of Fig.~\ref{fig:EXP_setup}, in the Supplementary Material (SM) of this paper and in the Supplementary Information of Ref. \onlinecite{Plati2024}).

To characterize the geometrical properties of the mixture, we define the size ratio $q=\sigma_S/\sigma_L$, the fraction of small grains $x_S=N_S/(N_S+N_L)$ and the area fraction $\phi=(N_S\sigma_S^2+N_L\sigma_L^2)\pi/4L^2$. 
Once a granular mixture with given geometrical properties $\{q,x_S,\phi\}$ is prepared, the experimental system is initialized by pouring the mixture of grains into the box and randomizing the initial configuration by spreading them manually. 
Then, the box is vertically accelerated following a prescribed sinusoidal signal $a_z(t)=g\Gamma\sin(2\pi ft)$. Vibrations are then characterized by the adimensional maximum acceleration $\Gamma$ and the frequency $f$ ($g$ is the gravity acceleration). In the SM, we provide more details about the tuning of the driving signals.

\begin{figure}
    \includegraphics[width=0.99\columnwidth,clip=true]{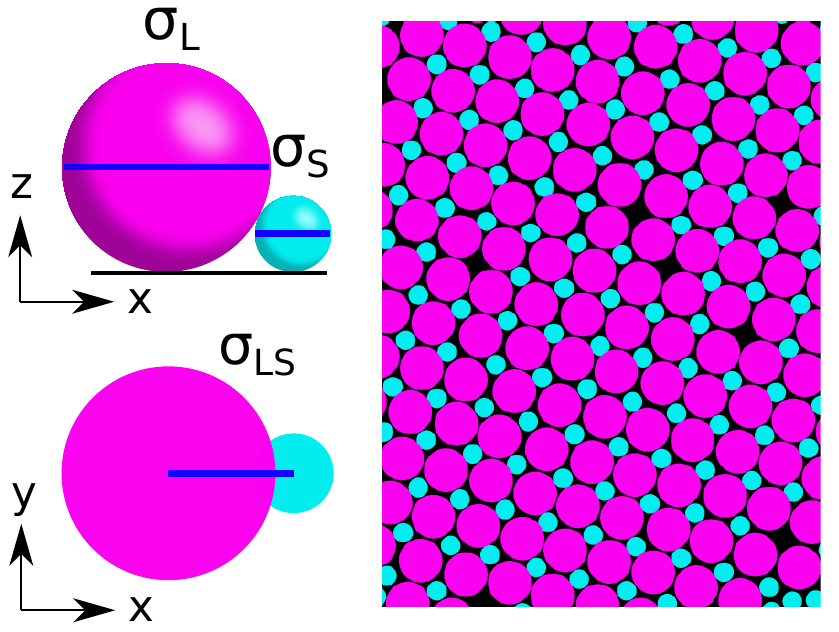}
    \centering
    \caption{Sketch of the equilibrium reference system of non-additive hard disks in which the S1 phase was first observed \cite{Fayen2020,Fayen2022}. On the left-hand side, we show the mapping between spheres laying on a plane and non-additive disks which undergo elastic collision as they reach a distance $\sigma_{LS}=\sqrt{\sigma_L\sigma_S}$. On the right-hand side, we show a portion of the S1 crystal assembled in the equilibrium reference system.} 
    \label{fig:EXP_q4all}
\end{figure}

Our granular mixtures have been prepared in a range of $q$, $x_S$ and $\phi$ for which a geometrically equivalent system undergoing thermal motion exhibits the spontaneous formation of a binary crystalline phase with a squared symmetry \cite{Fayen2020,Fayen2022}.
In Fig~\ref{fig:EXP_q4all}, we provide an essential sketch of this equilibrium reference system which consists of a binary mixture of non-additive hard disks effectively describing a system of spheres laying on a substrate. We also show a portion of the crystal self-assembled under thermal motion. For the rest of the paper, we will refer to this type of binary square crystal as S1 \cite{likos1993complex,Fayen2020,Fayen2022}. The adimensional values $\{q,x_S,\phi\}$, are referred to as the equilibrium or geometrical parameters. In contrast, the physical properties of the system that are not present in the hard-disk reference system (i.e. grain material, driving signals, roughness of the bottom plate, etc.) are designated as non-equilibrium parameters. 
In order to narrow down our parameter space, we first calibrated our driving conditions ($\{\Gamma,f\}$) by performing short self-assembly experiments with different drivings and optimizing for conditions where horizontal movement of the grains is fast, but small particles avoid stacking on top of each other. 
The best results were obtained with sinusoidal shaking at $\Gamma=1.79$ and $f=53$ Hz. Hence, we kept these values fixed for all other experimental runs.

\subsection{Self-assembly of the square binary crystal}

\begin{figure*}
    \includegraphics[width=0.99\textwidth,clip=true]{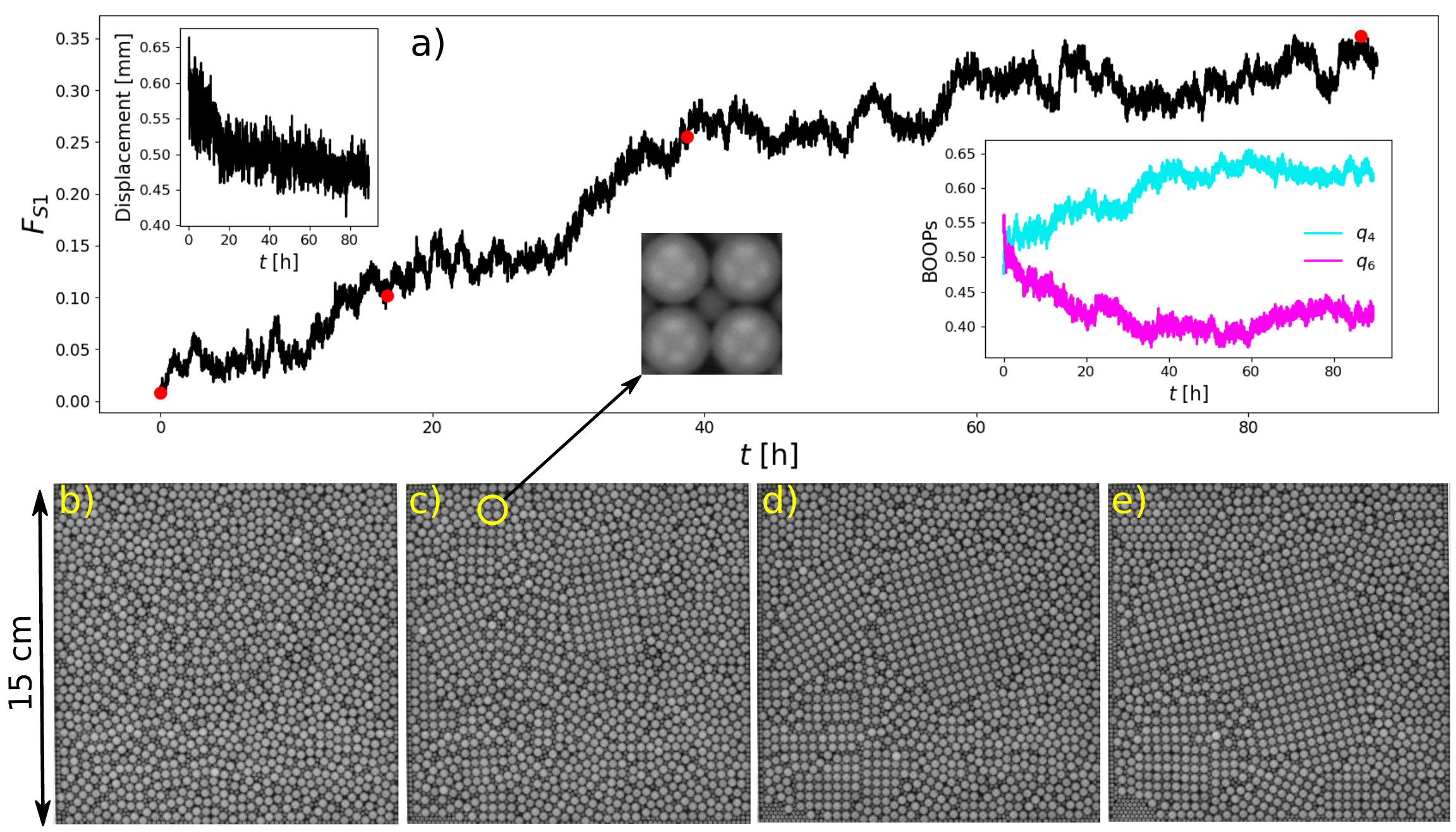}
    \centering
    \caption{Results from a long experiment with polyamide beads on a rough surface under sinusoidal vibration. Here $q=0.5$, $x_S=0.51$, $\phi=0.858$, $\Gamma=1.79$, $f=53$ Hz. a) Time evolution of the fraction of particles belonging to an S1 environment (see the text for the definition). The red dots mark the time at which the snapshots of the system shown in panels b-e) are taken (they are in chronological order). In the upper-left inset of panel a), we show the time evolution of the average large particle displacement over a time interval of 30 seconds (smoothed using a running average with a time window of 5 minutes). We use this quantity to characterize the average mobility of the system. In the bottom-right inset of panel a), we plot the time evolution of $q_4$ and $q_6$. Also shown here is a zoom-in on the unit cell of the S1 phase, which consists of four large grains surrounding a small one. } \label{fig:EXP_Photo_vs_time}
\end{figure*}

The first goal of the experiment was to investigate whether it is possible to obtain the S1 phase at the macroscopic scale using non-equilibrium athermal driving (i.e. mechanical vibration). In the equilibrium reference system, S1 self-assembly has been observed for $q\in[0.45,0.5]$, $x_S\in [0.45,0.6]$ and $\phi\geq 0.84$~\cite{Fayen2022}. To remain in the same range of geometrical parameters, we prepared a mixture of polyamide beads with $q=0.5$, $x_S=0.51$, $\phi=0.858$ and then performed a very long experiment ($\sim 90$ hours) with the previously found optimal driving sinusoidal driving ($\Gamma=1.79$, $f=53$ Hz).  Since size segregation phenomena are ubiquitous in granular matter \cite{Aumaitre2001,Kudrolli2004,Baldassarri2015,Aaranson2006,Rivas_2011,Huerta2004,WILLIAMS1976,AHMAD1973,Melby2007}, we expect the self-assembly of S1 crystals (where the two species are completely mixed) to compete with the formation of hexagonal domains of small and large particles.
In view of this, we characterize the evolution of our system with the square ($q_4$) and hexagonal ($q_6$) bond orientational order parameters (BOOPs). These are scalar observables accounting for square and hexagonal crystallinity defined as: 
\begin{equation}\label{eq::q46}
    q_{k}=\frac{1}{N_L}\sum_{m=1}^{N_L} q^m_{k}, \quad q^m_{k}=\frac{1}{N_m}\Big|\sum_{j=1}^{N_m} e^{ik\theta_{mj}}\Big|
\end{equation}
with $k=\{4,6\}$ and where $N_m$ is the number of nearest neighbours (identified through a cutoff distance) of the $m$th grain and $\theta_{mj}$ denotes the angle of the vector joining the centers of two particles with respect to a reference direction. In Eq.~\eqref{eq::q46}, we consider only large particles as these are sufficient for the characterization of the local orientational order.
To focus specifically on the S1 crystalline phase, we also introduce a different scalar observable namely the fraction $F_{S1}$ of large particles belonging to an S1 environment. This is obtained by counting the particles with  $q^m_4 > q_4^\star$ having at least three nearest neighbours with $q^m_4 > q_4^\star$, where $q_4^\star$ is an arbitrary threshold that we chose equal to 0.8. $F_{S1}$ is more precise than the global $q_4$ as an indicator of crystallization since it does not take into account the contribution of isolated particles with a high $q^m_4$ (which can also be present in a fluid phase). In Fig.~\ref{fig:EXP_Photo_vs_time}a, we plot $F_{S1}$ as a function of time and show a zoom of an S1 unit cell formed in the experiment; in Fig.~\ref{fig:EXP_Photo_vs_time}b-e, we provide the direct visualization of the self-assembly process at four consecutive times.
At the beginning, the system is completely amorphous with $F_{S1}\sim0$ (Fig.~\ref{fig:EXP_Photo_vs_time}b), then, for the first 40 hours, we observe a time evolution of $F_{S1}$ consisting of long stationary periods interrupted by relatively fast increases which correspond to the growth of large S1 clusters with different orientations (Fig.~\ref{fig:EXP_Photo_vs_time}c and \ref{fig:EXP_Photo_vs_time}d). At the same time, the BOOPs plotted in the bottom right inset show an increase in the average local square orientational order and a decrease in the hexagonal one. We point out that the main S1 formation is observed around the few largest clusters nucleated at relatively short times. 
The remaining time of the experiment is characterized by a very slow increase of $F_{S1}$ during which the two largest clusters coarsen into a single S1 domain spanning a macroscopic fraction of the system (Fig.~\ref{fig:EXP_Photo_vs_time}e). We also observe the formation and growth of smaller S1 clusters in the bottom part of the system. As expected, the self-assembly process is accompanied by a gradual decrease in particle mobility, which we define as the average absolute displacement of large particles over a time interval of 30 seconds (upper-left inset of Fig.~\ref{fig:EXP_Photo_vs_time}a).  
During the second half of the experiment, we note the appearance of hexagonal domains of small and large grains in different parts of the system. This is reflected also in the time evolution of the BOOPs where we see an increase of $q_6$ starting around $t=60$ h. We can explain this effect by considering that the S1 structure requires a homogeneous composition of small and large grains; therefore, starting from a local amorphous environment with a non-perfectly homogeneous composition, the formation of S1 will tend to expel the excess species. This in turn can lead to local regions with a high concentration of one of the species, which favours the formation of hexagonal domains. On top of that, over very long experiments the horizontal calibration of the system may slowly get lost, introducing the effect of gravity and inhomogeneity in the granular dynamics. This promotes size segregation \cite{WILLIAMS1976,AHMAD1973,Huerta2004} and may therefore play a role in the formation of hexagonal domains. However, the important thing to note here is that these effects are rather marginal compared to the mixing of the two species within the S1 structure which continues to grow despite the formation of competing structures.  
The reproducibility of this result is supported by a shorter experiment (30 hours) with a slightly lower area fraction ($\phi=0.851$) that exhibited the formation of S1 domains in the early dynamics (see SM).

Finally, we remark that mixing particles of different sizes in a granular system is a paradigmatic problem particularly relevant in different applications \cite{BRIDGWATER2012,Hui2024,WILLIAMS1976}. As argued in a recent numerical study \cite{nakai2024}, the self-assembly of binary crystals in a polydispersed granular system can be a way to address the problem of size segregation \cite{Aumaitre2001,Rivas_2011,Melby2007,WILLIAMS1976}. Our experiments suggest that mixing of particles by athermal self-assembly may indeed be feasible in a realistic granular system. 
In order to test the robustness of this self-assembly behavior, we repeated our experiments with steel beads (see SM), and found a qualitatively similar behavior. In particular, steel beads can also self-assemble into large domains of S1 crystal, but the overall crystallinity of the system typically remains lower.

\subsection{Liquid-crystal phase coexistence}

\begin{figure*}
\includegraphics[width=0.99\textwidth,clip=true]{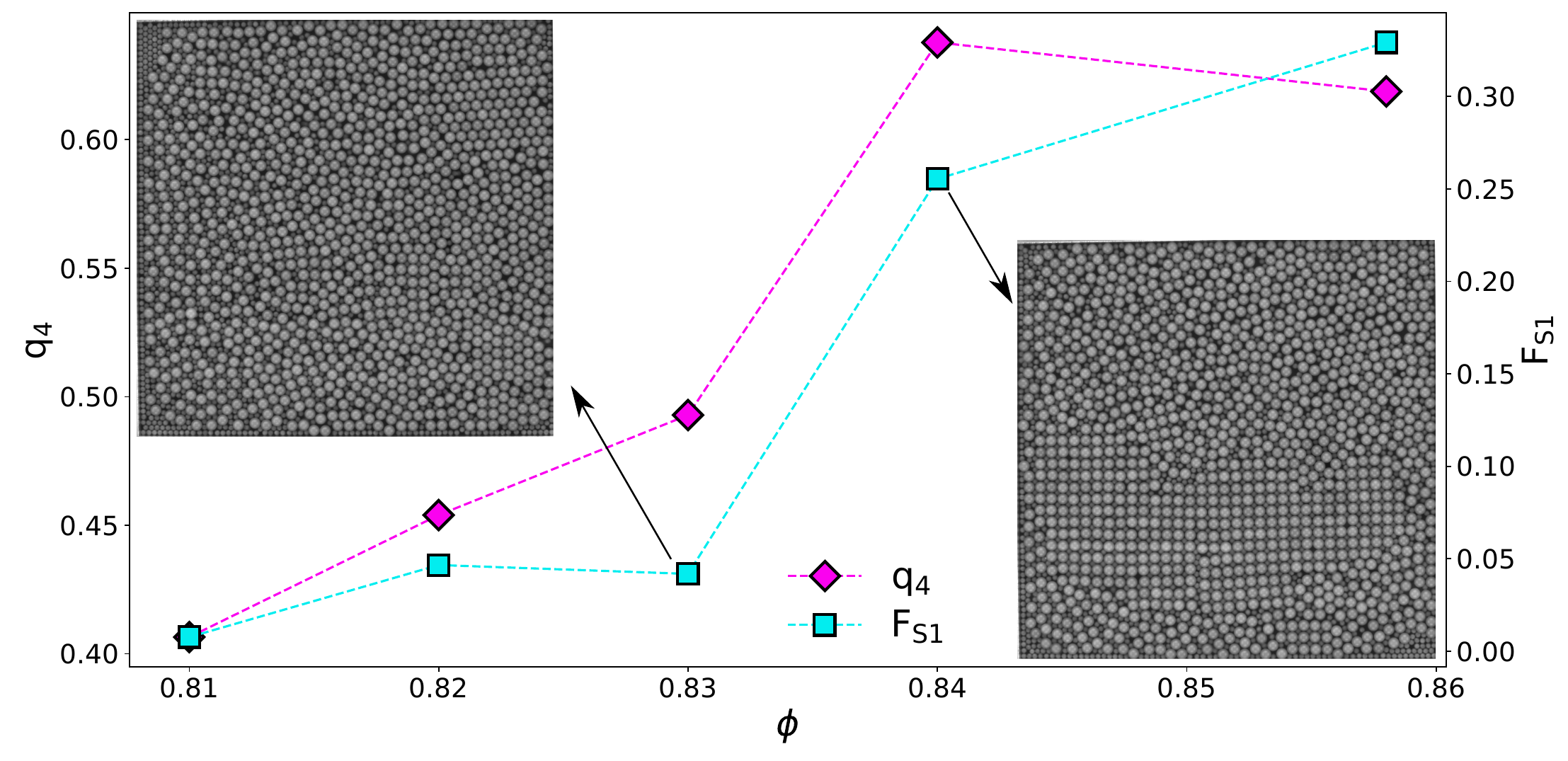}
    \centering
    \caption{Time averaged $q_4$ and $F_{S1}$ as a function of the area fraction for experiments with polyamide beads on a rough surface under sinusoidal vibrations. In all the cases $q=0.5$, $x_S=0.51$, $\Gamma=1.79$ and $f=53$ Hz. Time averages are performed over the last 50 minutes of the experiments. The total duration $t_{tot}$ varies for different experiments, we have $t_{tot}=\{20,20,20,29,89\}$ hours corresponding to the explored area fraction $\phi=\{0.81,0.82,0.83,0.84,0.858\}$. In the two insets, we provide the final snapshots for the experiments performed at $\phi=0.83$ (top left) and $\phi=0.84$ (bottom right).} 
    \label{fig:EXP_Photo_vs_phi}
\end{figure*}

In order to explore how the area fraction affects crystallization behavior, we perform a series of experiments at lower area fractions down to $\phi=0.81$, while keeping the other parameters fixed. In Fig.~\ref{fig:EXP_Photo_vs_phi}, we show $q_4$ and $F_{S1}$ averaged over the last 50 minutes of each experiment as a function of $\phi$ (the total duration $t_{tot}$ varies for different experiments, see caption). We note a sharp increase of both observables between $\phi=0.83$ and $\phi=0.84$. The snapshots of the final configurations obtained in these cases (shown as insets) suggest that the system undergoes a liquid-solid-like phase transition as a function of the area fraction. For $\phi=0.83$ the system exhibits an amorphous structure with no significant formation of S1 domains and the same holds also for lower $\phi$ (not shown). For $\phi=0.84$, we found a single large domain of S1 particles coexisting with the amorphous phase. Finally, the last point at $\phi=0.858$ coincides with the experiment discussed in the previous section, its final configuration is the one shown in Fig.~\ref{fig:EXP_Photo_vs_time}e where the S1 domains are widespread all around the system.
By looking at the final configurations in Fig.~\ref{fig:EXP_Photo_vs_phi}, it is possible to note a gradient of decreasing concentration of large particles from right to left (the same is also observed at $\phi<0.83$). Of course, this does not happen in the S1 domain at $\phi=0.84$ where the two species are perfectly mixed. We argue that this is a gravity-induced size segregation effect \cite{WILLIAMS1976,AHMAD1973,Huerta2004} due to the loss of horizontal calibration in the system. This was not observed in previous experiments done with the same setup \cite{Plati2024}, likely because they were performed at much lower vibration intensity. However, this experimental imperfection turns out to highlight the robustness of the S1 self-assembly process when the right geometrical conditions are matched. Indeed, S1 domains are able to homogenize the particle concentration even when gravity would promote size segregation. The systematic characterization of the competition between size segregation and self-assembly in granular binary mixtures goes beyond the scope of this paper and is left for future studies. 
On a practical level, this effect provides us with a qualitative criterion for deciding when to stop the experiments. If, after 20 hours of vibration, S1 self-assembly is not observed and relevant size segregation is found (with the consecutive formation of hexagonal clusters of small and/or large particles), we consider the experiment terminated. Otherwise, we let the experiment run until the formed S1 domains stop growing significantly.

\begin{figure}
    \includegraphics[width=0.49\columnwidth,clip=true]{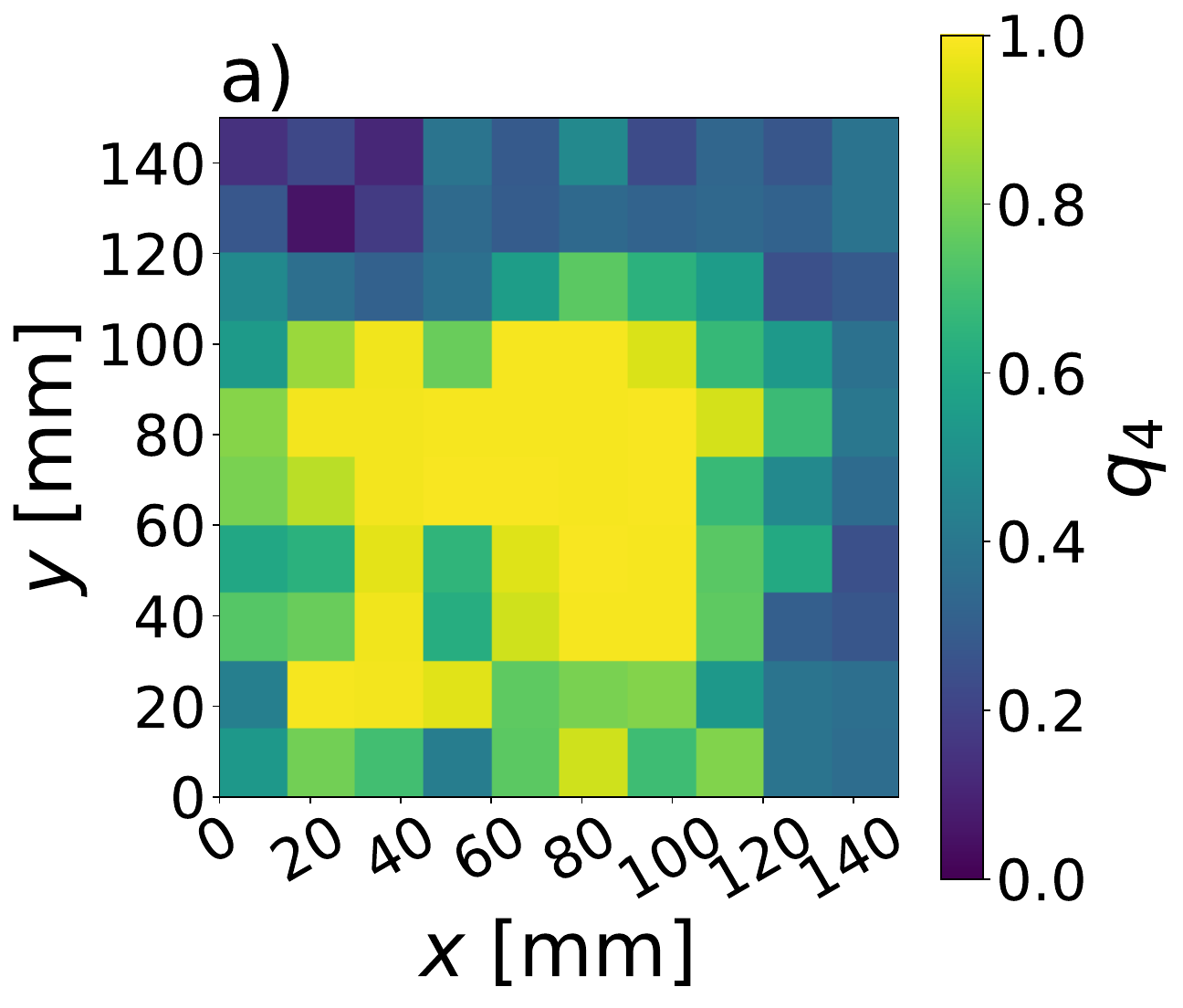}
    \includegraphics[width=0.49\columnwidth,clip=true]{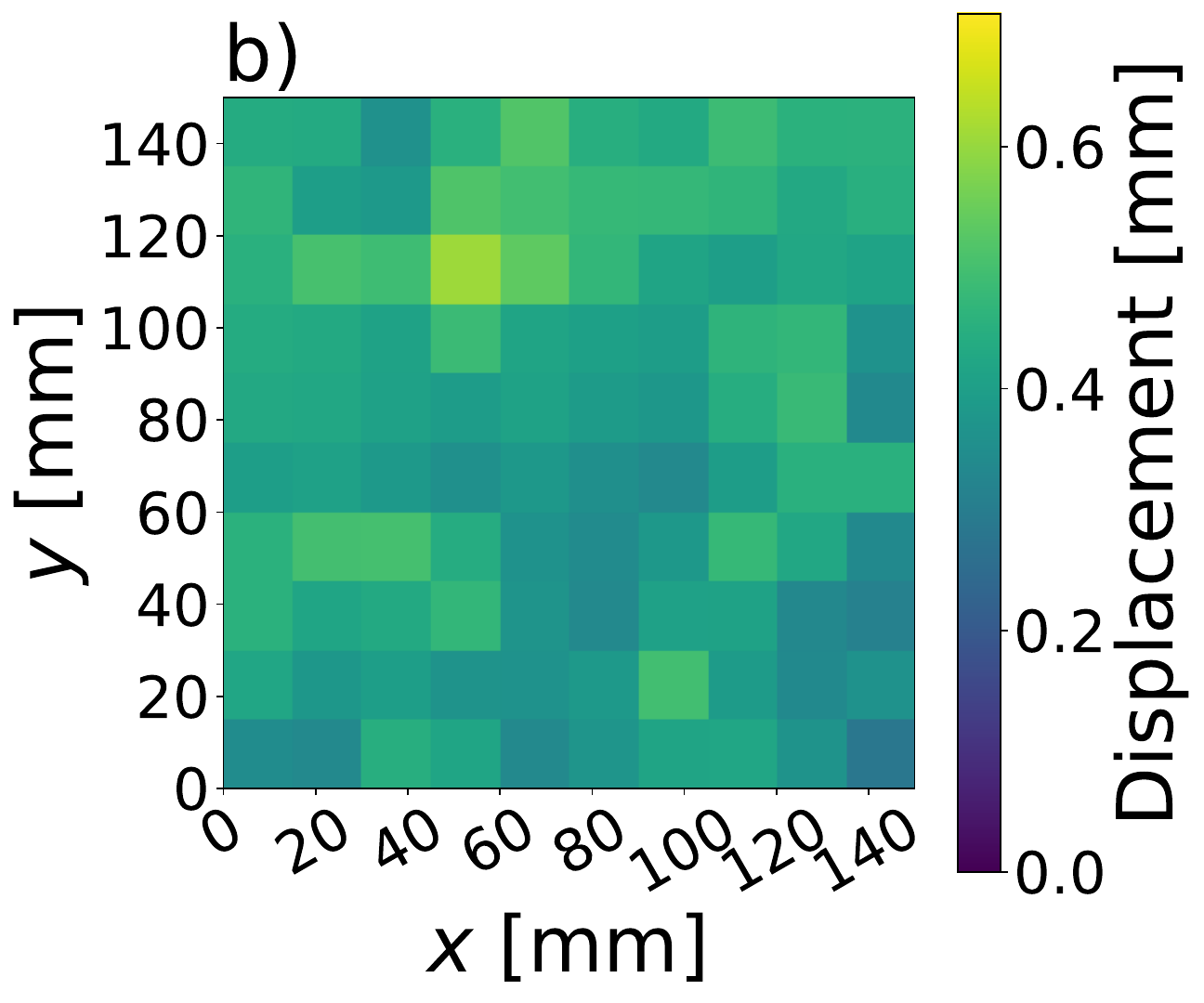}
    \includegraphics[width=0.49\columnwidth,clip=true]{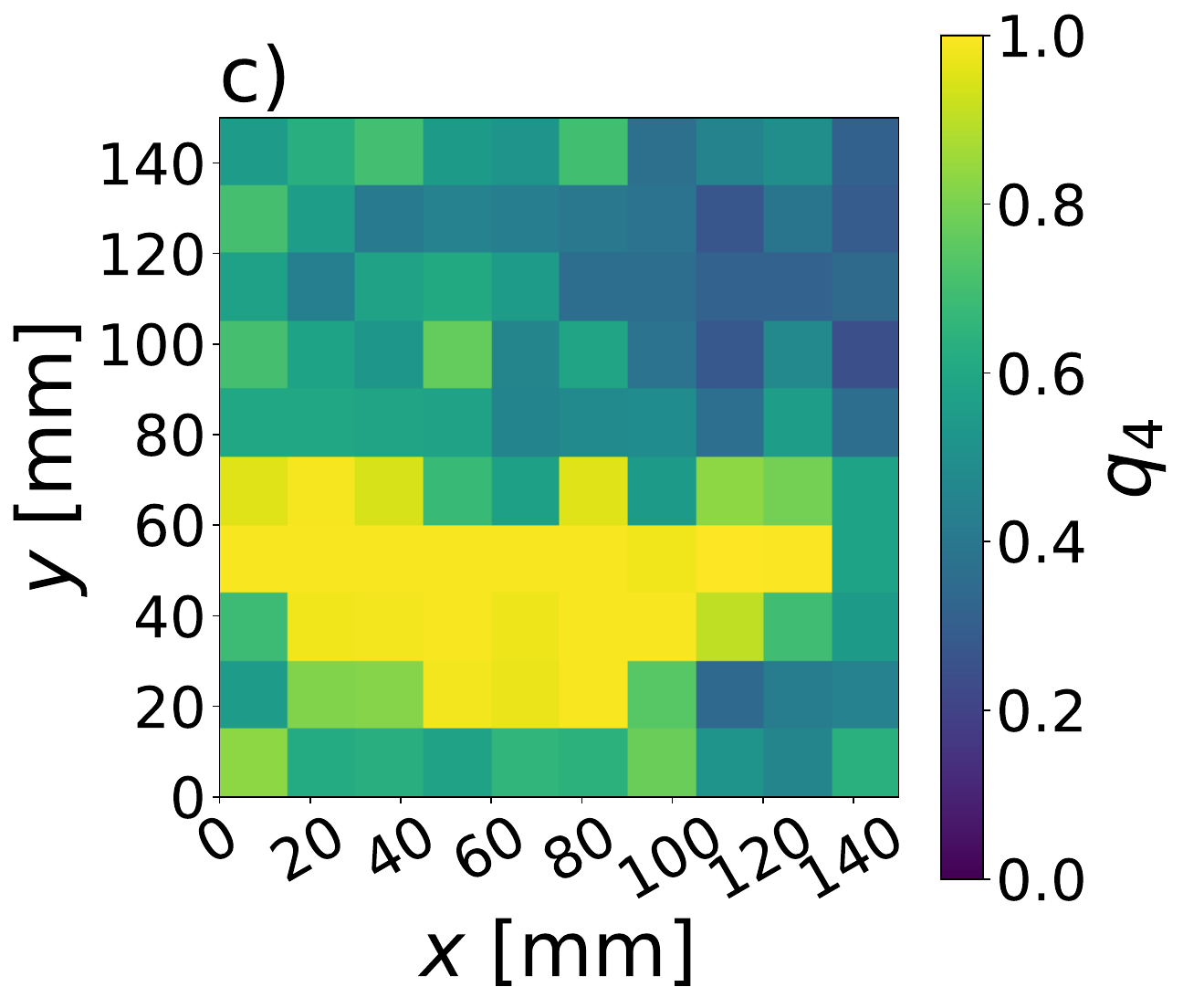}
    \includegraphics[width=0.49\columnwidth,clip=true]{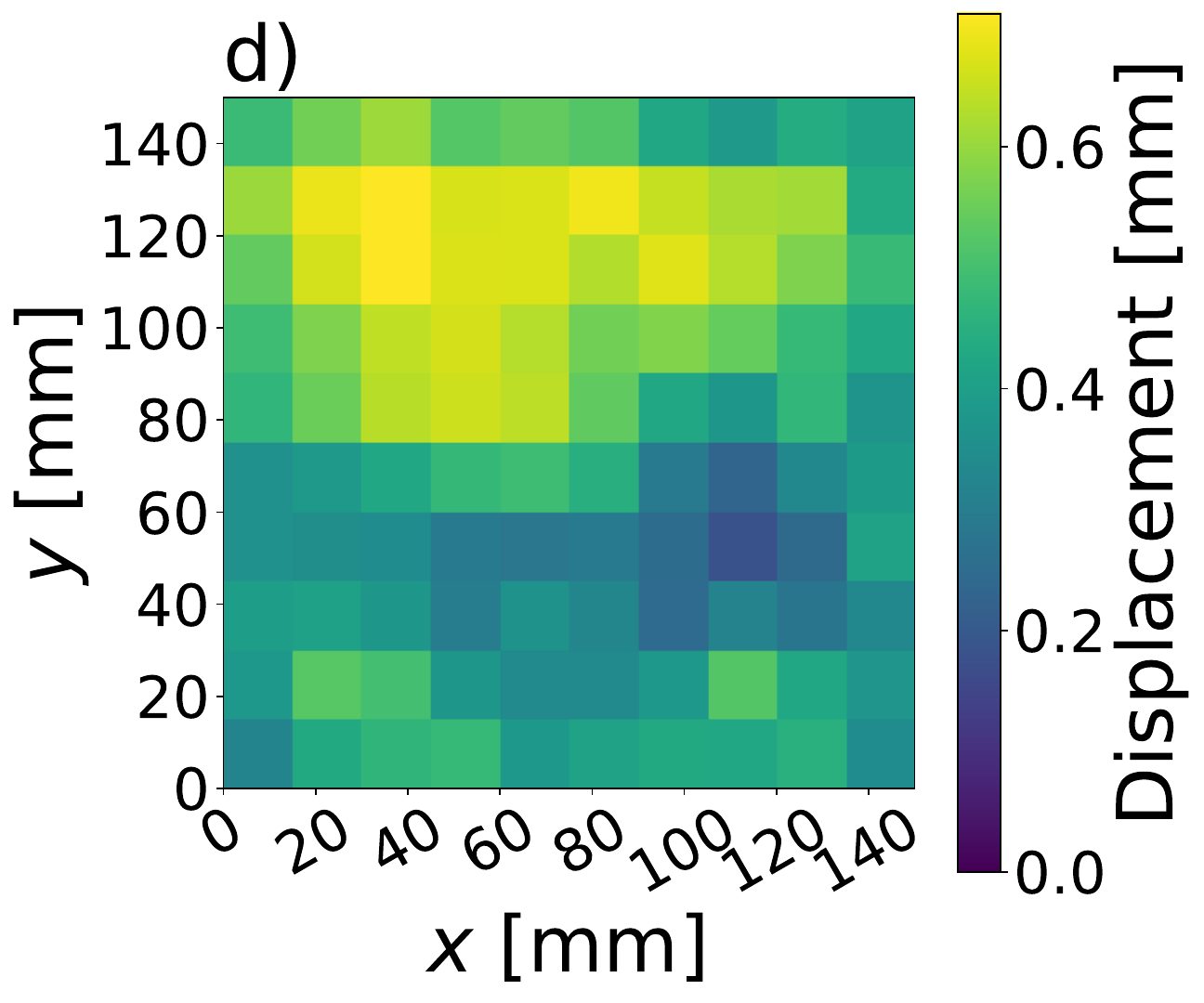}
    
     \caption{Comparison between the maps of average $q_4$ and large particle displacement over 0.17 seconds
     for the experiments performed at $\phi=0.858$ (panel a and b) and $\phi=0.84$ (panel c and d).  
     In the first case, the displacement field is homogeneous regardless of the local value of $q_4$, in the latter, low mobility regions correspond to the high $q_4$ cluster. The analysis is performed over a 60-second-long high-rate acquisition (60 fps) done at the end of the experiments. Each shown map results from an average of 3520 single-frame analyses (or double-frame in the case of the displacement). } 
    \label{fig:Exp_q4VSdisp}
\end{figure}
The formation of a single crystal domain at $\phi = 0.84$ is strongly reminiscent of an equilibrium phase coexistence between a fluid and a crystal. This picture is further reinforced by the observation that the fluid phase in this system remains highly mobile, ruling out the possibility that the crystal has stopped growing because the system reached dynamical arrest. To demonstrate this, we compare in Fig.~\ref{fig:Exp_q4VSdisp} the spatial variation in the averaged local $q_4$ with the one of large particle mobility for $\phi=0.84$ and $\phi=0.858$. Each map is performed by dividing the system into 15 mm side square subregions and averaging quantities over a 60-second-long high-rate acquisition (60 fps) done at the end of the experiments. To compute the local $q_4$, we simply average $q^m_4$ over each subregion, while as a measure of local mobility, we use the mean absolute displacement of the grains over a time interval of 0.17 seconds. From these maps, it is clear that, for $\phi=0.858$, the mobility is quite homogeneous all over the system regardless of the local $q_4$. This means that the S1 phase is coexisting with other solid- or glassy domains (like the hexagonal one in the top-right corner of Fig.~\ref{fig:EXP_Photo_vs_time}e). .  
In contrast, for $\phi=0.84$, the displacement map exhibits low mobility in the S1 phase (i.e. the region of large local $q_4$) and a much higher one in the surrounding fluid. The videos of the two experiments we provide in the SM further clarify the exposed picture. 
We point out that this phenomenology is consistent with the interpretation of the system at $\phi=0.84$ forming a stable coexistence between a mobile fluid phase and a solid S1 cluster. We confirmed the reproducibility of this behavior in two additional runs (not shown) at $\phi=0.84$, where we again observed the formation of low-mobile S1 clusters in coexistence a fluid.

In the remainder of the paper, we will further investigate the non-equilibrium phase coexistence between the granular binary crystal and the fluid phase found in our experiments by means of numerical simulations.

\section{Numerical Simulations} \label{sec: simu}

As discussed in the previous sections, experiments to study the self-assembly of the S1 phase are extremely long and particularly hard to calibrate. In order to further explore the self-assembly process and better characterize the non-equilibrium phase coexistence between the granular liquid and the S1 phase, we now turn our attention to numerical simulations.

First, we study the numerical reproduction of the experimental setup using a realistic granular model that can be simulated through molecular dynamics based on the Discrete Element Method (DEM) \cite{Cundall79,PoeschelBook}. This analysis aims to show that our numerical simulations reproduce the S1 self-assembly process observed in the experiments.

Once the numerical model has been validated, we move to the implementation of the direct phase coexistence method \cite{Opitz74,Ladd77,Ladd78,Cape78,Espinosa2013, Smallenburg2024} to systematically explore the behavior in the coexistence region. This will be done both with DEM simulations of the realistic granular model and with Event Driven Molecular Dynamics (EDMD) \cite{Smallenburg2022} of a simplified model which is able to capture the minimal ingredients of the granular dynamics under study~\cite{Plati2024, Maire2024interplay}. The direct phase coexistence method will allow us to identify the range of area fractions for which the liquid and the S1 phases are found in stable coexistence at different densities and, remarkably, at different granular temperatures.

\begin{figure*}
\includegraphics[width=0.9\textwidth,clip=true]{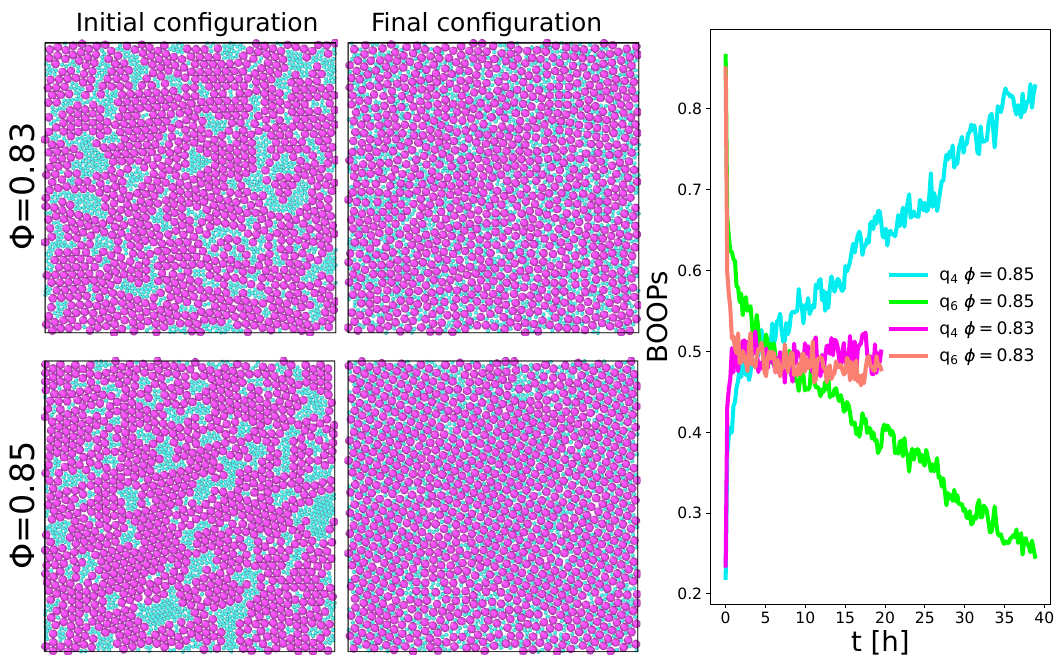}
    \centering
    \caption{DEM simulations of the quasi-2D vibrated setup with PBC.
    On the left, we show the
initial and final configurations for $\phi=0.83$ and $\phi=0.85$ which evolve respectively towards a fluid and an S1 phase. On the right, we plot the BOOPs as a function of time for the same simulations. For all cases we use $N=2000$, $q=0.476$, $x_S=0.5$, $\Gamma=4.4$ and $f = 350$ Hz.} 
    \label{fig:NUM_SA_periodic}
\end{figure*}

\subsection{In silico replication of the experimental setup} \label{sec:insilicoRep}
DEM simulations consist of molecular dynamics that make use of realistic contact models to describe the motion of the grains during collisions with other grains or confining walls. This allows the reconstruction in silico of granular experimental setups, taking into account their real-world sizes and material properties. In our case, we use the Hertz-Mindlin model \cite{LammpsSiteGran,Zhang2005,PoeschelBook} for the contact dynamics as implemented in LAMMPS \cite{Plimpton1995,Plimpton2022,LammpsSiteGran}. This model contains different parameters referring to the elastic and dissipative contribution of normal and tangential forces between colliding grains. The parameters related to elastic interactions are directly linked to the material properties (i.e. Young modulus and Poisson ratio) while the ones needed to model the dissipative terms must be fixed according to a calibration procedure. In our simulations, all the model parameters are tuned according to previous studies \cite{PlatiSlow2020,Plimpton2022} which revealed a good qualitative agreement between experiments and simulations. We point out that we always use a Young's modulus lower than that of the real materials; this is a compromise that allows us to use a larger discretisation time step for the simulations thus reducing their computational cost.        
Our numerical setup mimics the experimental one: It is contained in a simulation box of height $h$ and width $L \gg h$ vertically confined by two horizontal plates. For the horizontal directions, we consider both Periodic Boundary Conditions (PBC) and hard walls. External energy is provided to the grains by imposing a sinusoidal vertical acceleration $a_z(t)=g\Gamma\sin(2\pi ft)$ to the boundaries of the system. In a preliminary analysis, we found that for the numerical system, a good driving condition to prevent grains from exploring the vertical direction when vibrated is $f=350$ Hz and $\Gamma=4.4$. It reasonably differs from the one used in the experiments since the simulated grains have material properties that do not perfectly match the real ones. Moreover, in this frequency range, we observed that the experimental system is subjected to the effect of vibrational modes of the bottom plate. Obviously, this does not occur in the simulation where hard boundaries are perfectly rigid. More details on the implementation of our DEM simulations can be found in the Supplementary Information of Ref.~\onlinecite{Plati2024}. 

\begin{figure}

    \includegraphics[width=0.99\columnwidth,clip=true]{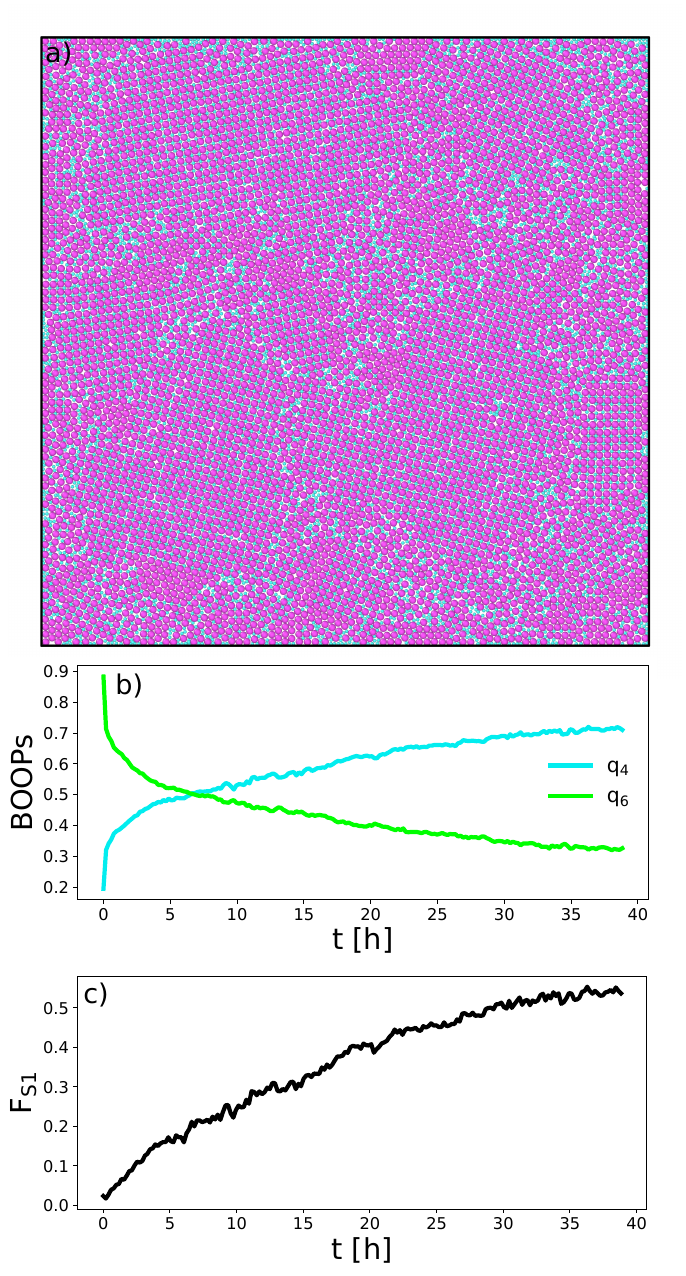}
    \centering
    \caption{DEM simulation of the quasi-2D vibrated setup with hard walls for a large system size $N=12000$.  a) Final configuration, b) time evolution of BOOPs and c)fraction of particles belonging to an S1 environment. The geometrical parameters are $\phi=0.85$, $q=0.476$, $x_S=0.5$; the driving ones are $\Gamma=4.4$ and $f=350$ Hz.} 
    \label{fig:DEM_Snap}
\end{figure}

We initialize our system with a granular mixture having the desired geometrical parameters $\{q,x_s,\phi\}$ and random horizontal velocities extracted from a Gaussian distribution.
In Fig.~\ref{fig:NUM_SA_periodic}, we show the initial and final configurations as well as the evolution of the BOOPs for two simulations with PBC at different area fractions $\phi=83$ and $\phi=0.85$.  
The system is prepared in an initial configuration where large and small grains are segregated into single-component hexagonal patches. For $\phi=0.85$, once the vibration is turned on, the hexagonal domains melt and the two-grain species mix forming the granular S1 phase. Instead, at $\phi=0.83$, the system evolves towards a homogeneous fluid phase. 
To get even closer to realistic conditions and test the robustness of the self-assembly process with respect to the system size, we now consider a larger system confined by hard walls at $\phi=0.85$. In Fig.~\ref{fig:DEM_Snap}, we show the related numerical results for the final configuration, the BOOPs and the fraction of particles belonging to an S1 environment. First, we note that the formation of the crystalline phase occurs also in this case.  Interestingly, the self-assembly of S1 in the simulations seems to be noticeably faster than in the experiment. If we compare the best experimental run in Fig.~\ref{fig:EXP_Photo_vs_time} with the two numerical ones at $\phi=0.85$ in Fig.~\ref{fig:NUM_SA_periodic} and \ref{fig:DEM_Snap}, we see that in the latter, the crystallization process is still ongoing at the end of the simulation even if a macroscopic fraction of the system is already in the S1 phase. 
We point out that in the simulations, hexagonal domains seem much easier to melt and this can depend on several causes such as the fact that the simulation box is perfectly horizontal, the different driving parameters or the specific granular interaction forces used in the numerical model. The latter are realistic but still coarse-grained, and hence cannot capture all the possible effects that are present in the experiments.

To summarize, this numerical analysis confirms that DEM modeling of granular materials can qualitatively reproduce the S1 self-assembly found in the experimental setup. This also means that this phenomenon is general, rather than a consequence of some peculiarities of our experimental setup. 

\subsection{Direct phase coexistence}
In this section, we numerically explore the behavior of the system in the coexistence region between the granular liquid and the S1 crystal using the direct phase coexistence method implemented both with DEM and EDMD simulations.

In thermal systems, a common approach for determining equilibrium phase boundaries between coexisting phases involves performing simulations in which the two phases are in coexistence with each other. The ability of the two phases to exchange energy, volume, and particles then naturally causes the two phases to evolve towards thermal, mechanical, and chemical equilibrium.  This is the main idea behind the direct phase coexistence method \cite{Opitz74,Ladd77,Ladd78,Cape78}. In this approach, one usually puts two slabs of the different phases into contact in a periodic simulation box elongated in one direction. This geometrical setting usually stabilizes the interface between the two phases if stable coexistence is reached. Depending on whether the global value of the control parameter imposed on the system is inside or outside the coexistence range, the system can either equilibrate towards a single phase or towards stable coexistence. For coexistences involving crystals, special care needs to be taken in order to ensure that the crystal is not under any artificial strain imposed by the boundaries of the simulation box \cite{Espinosa2013, Smallenburg2024}. One crucial advantage of the direct coexistence method is the fact that its implementation does not require an assumption of thermodynamic equilibrium (in contrast to e.g. free energy calculations \cite{Frenkel84}), and as a result, it has also been applied to non-equilibrium systems such as active matter \cite{Mandal2019, van2020predicting}. 
\begin{figure}
    \includegraphics[width=0.85\columnwidth,clip=true]{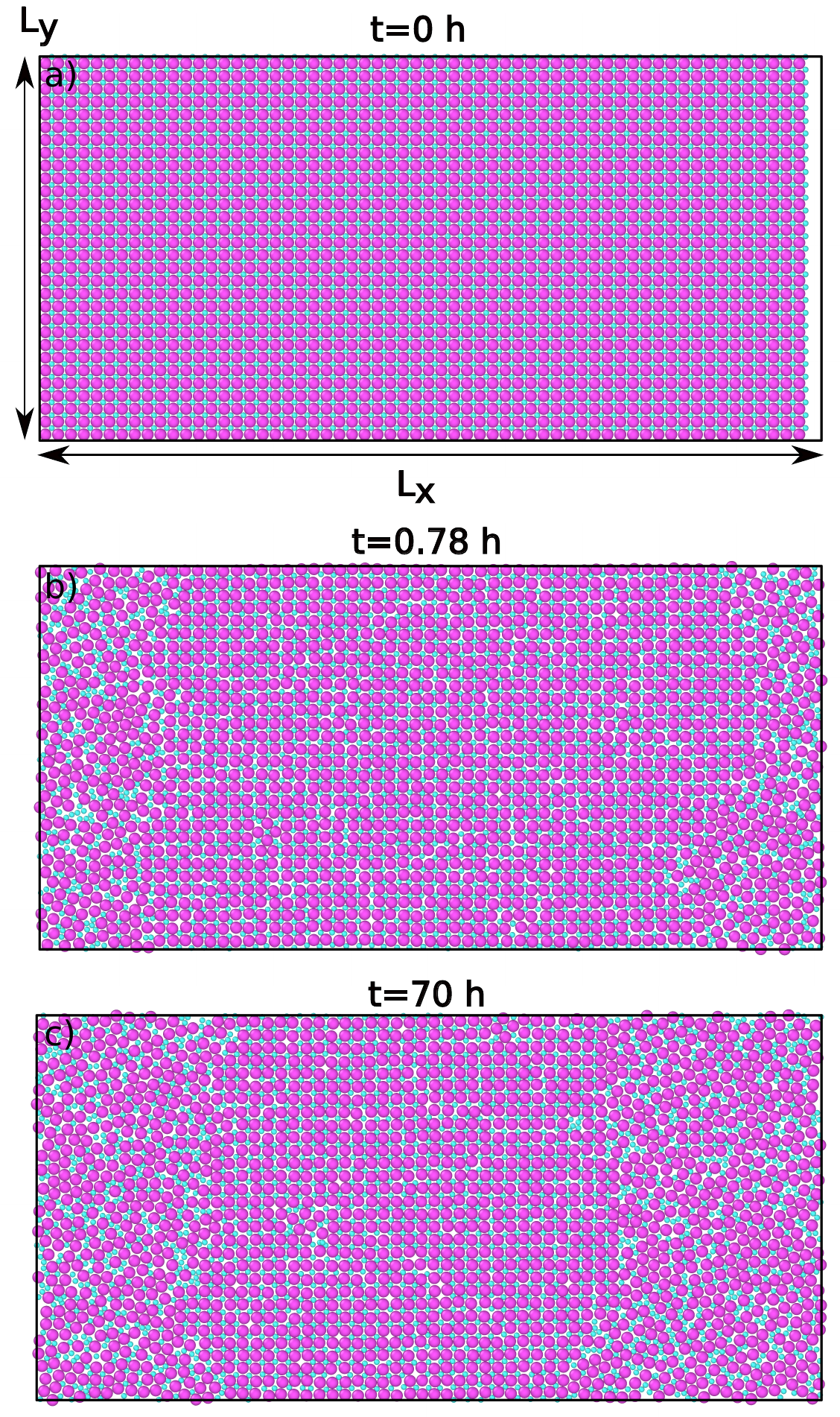}
    \includegraphics[width=0.85\columnwidth,clip=true]{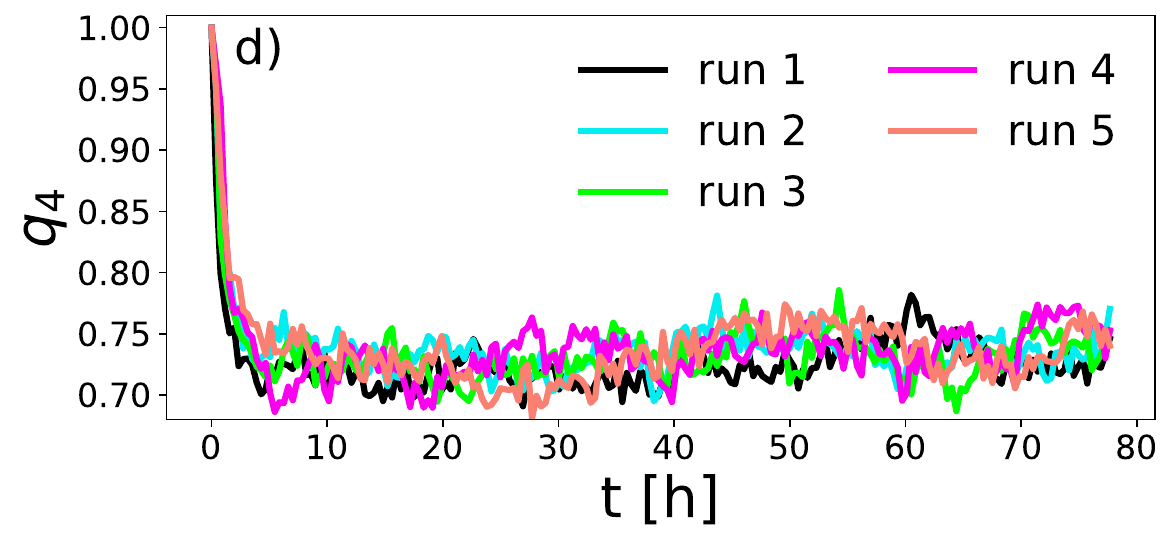}
    \centering
    \caption{Direct phase coexistence method for the DEM model. We show snapshots a) of the initial configuration, b) of a transient state, and c) of the non-equilibrium stationary state. We note that in the initial state, there is some free space between the right edge of the initial crystalline slab and the edge of the simulation box. In panel d), we plot the average $q_4$ as a function of time for five statistically independent runs. Simulations were performed with $N=3600$, $\phi=0.84$, $q=0.476$, $x_S=0.5$, $\Gamma=4.4$ and $f=350$ Hz.} 
    \label{fig:DEM_SnapCoex}
\end{figure}
\subsubsection{DEM Simulations}

Our DEM numerical setup is the same as the one used in Sec.~\ref{sec:insilicoRep} but now the dimensions of the simulation box are $L_x \times L_y \times h$ with $L_x=\lambda L_y$ where $\lambda>1$. The number $N_L=N_S=1800$ and diameters $\sigma_L=2.5$ mm, $\sigma_S=0.476\sigma_L$ of the grains are fixed while $L_y$ and $\lambda$ are chosen to satisfy the desired global area fraction $\phi$. Instead of initializing the system with a liquid and a crystalline slab in contact, we prepare it with a single crystalline block in the left-hand region of the box. The initial lattice spacing $a_y$ of this S1 domain is such that it occupies all the space in the $y$-direction. Then, depending on the global $\phi$ and the chosen area fraction of the crystal slab, there will be a certain amount of free space between the right edge of the initial crystalline block and the one of the simulation box (see Fig.~\ref{fig:DEM_SnapCoex}).
Once initial velocities are randomly set according to a Gaussian distribution and vertical vibrations are switched on, the grains start to explore the available space. If $\phi$ belongs to the 
phase coexistence region, we expect from the equilibrium phenomenology that a part of the crystal will melt and the system will end up in a stable coexistence between a crystalline and a liquid slab. Despite our system being far from thermodynamic equilibrium, these expectations remain satisfied as we show for a specific case at $\phi=0.84$ in Fig.~\ref{fig:DEM_SnapCoex}. From the plot of the $q_4$ as a function of time, we can see that, after a transient, the system reaches a non-equilibrium steady state with a stable degree of crystallinity. We remark that, in general, direct coexistence simulations lead to
a deformed crystal with a different lattice spacing along $x$ and $y$ \cite{Espinosa2013, Smallenburg2024}.  To avoid this, one has to fine-tune the initial value of $a_y$ (or equivalently, the area fraction of the initial crystal slab) such that it is equal to the final $a_x$. All the results presented in this section have been obtained with $a_y=1.0966\sigma_L$ corresponding to an initial crystal slab with area fraction 0.857. This gives the formation of an undeformed crystalline slab at coexistence (see the SM for our calibration procedure). 

\begin{figure}

\includegraphics[width=0.98\columnwidth,clip=true]{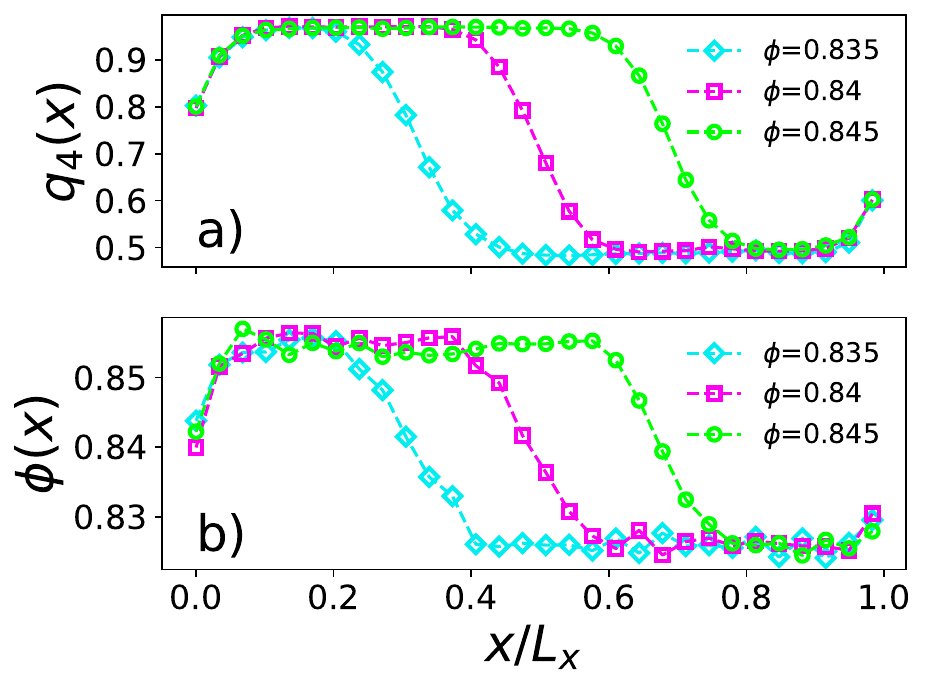}

    \centering
    \caption{Horizontal profiles of $q_4(x)$ and $\phi(x)$ obtained with the direct phase coexistence method implemented for the DEM model. We report the results for three different imposed global area fractions $\phi=\{0.835,0.84,0.845\}$ and $N=3600$, $q=0.476$, $x_S=0.5$, $\Gamma=4.4$, $f=350$ Hz.} 
    \label{fig:DEM_Profile}
\end{figure}

Since the system at $\phi=0.84$ exhibits a phenomenology coherent with the one expected in the coexistence region, we now focus on the profiles of $q_4$ and area fraction around this state point.  In Fig.~\ref{fig:DEM_Profile},  we plot the local BOOP $q_4(x)$ and local area fraction $\phi(x)$ along $x$ for $\phi=\{0.835,0.84,0.845\}$.  For a given simulation, local observables are calculated based on bins of size $L_y \times \Delta x$ with $  \Delta x=2.2\sigma_L$ placed at $x \in [i\Delta x, (i+1)\Delta x]$ and averaged over 150 snapshots separated by 0.2 h in the steady state. The obtained profiles are then averaged again over three independent simulations. For each couple of profiles at a given global $\phi$, we observe a denser region with area fraction $\phi_{sol}$ corresponding to the S1 slab and a less dense one at $\phi_{liq}$ corresponding to the liquid phase. We note that the portion of the system occupied by the crystal increases with the global area fraction $\phi$ while $\phi_{sol}$ and $\phi_{loc}$ do not depend on it. This suggests the existence of an underlying lever rule for the density even if the system is not described by a free energy. Such behavior was also found and theoretically explained in active systems \cite{evans2023theory,chiu2024theory, hermann2024active}. We can compute the average local area fraction in the two regions obtaining $\phi_{sol}=0.855 \pm 0.001$ and $\phi_{liq}=0.826 \pm 0.001$. To avoid the effect of the interface, $\phi_{sol}$ is calculated by averaging over bins with  $q_4>0.96$ while $\phi_{liq}$ is calculated by averaging over bins with $q_4^i<0.5$. 

From this analysis, we conclude that the coexistence region of this granular liquid-solid-like phase transition is $\phi \in [0.826,0.855]$. Remarkably, this is quantitatively coherent with the experimental results where we observed phase coexistence at $\phi=0.84$.

\subsubsection{EDMD Simulations}
DEM simulations are particularly suitable to provide a direct link between real experimental setups and numerical simulations. Now we change perspective and approach the problem of granular phase coexistence with a model that provides a direct link with the reference equilibrium system where the self-assembly of the S1 phase was observed~\cite{Fayen2022}. We consider the granular version of the non-additive hard-disk system described in Fig.~\ref{fig:EXP_q4all}a which is fully 2D and where collisions between particles are considered instantaneous events. The crucial differences in the granular case are the coefficient of restitution $\alpha \in [0,1]$ related to energy loss at collision and the explicit presence of a thermostat characterised by an inverse characteristic time  $\gamma$ and a temperature $T_b$ (which, in order to be consistent with granular temperatures, we write in units of energy). This is implemented by imposing noise $\sqrt{2\gamma T_b/m_i}\bm{\xi} (t)$ on the grains where $m_i$ is the mass of particle $i$ and $\bm{\xi}=\{\xi_x,\xi_y\}$ represents a 2D delta-correlated continuous random variable distributed according to a Gaussian with zero mean and unitary variance: $\langle \bm{\xi}(t) \rangle=0$, $\langle \xi_i(t)\xi_j(t') \rangle=\delta_{ij}\delta(t-t')$. To summarize, in between collisions particle dynamics evolve following the Langevin dynamics \cite{baldassarri2005temperature}
\begin{equation}\label{eq:langevin}
   \dot{\bm v}_i(t)=-\gamma \bm v_i(t) + \sqrt{2\gamma T_b/m_i} \bm{\xi}(t)
\end{equation}
while particle collisions are governed by
\begin{equation}
    \begin{split}
        \bm v_i'&= \bm v_i + \mu_{ji}(1+\alpha)(\bm v_{ij}\cdot \bm{\hat{\sigma}}_{ij})\bm{\hat{\sigma}}_{ij} \\
        \bm v_j'&= \bm v_j - \mu_{ij}(1+\alpha)(\bm v_{ij}\cdot \bm{\hat{\sigma}}_{ij})\bm{\hat{\sigma}}_{ij} ,
    \end{split}
    \label{eq:collrule}
\end{equation}
where $\mu_{ij}={m_i}/(m_i+m_j)$ and $\bm v_i'$ is the post-collisional velocities of particle $i$, while $\bm{\hat{\sigma}}_{ij}$ and $\bm v_{ij}$ are respectively the unit vector joining particles $i$ and $j$ and the relative velocity between them.
For numerical simulations, we used a hybrid time-stepped/EDMD scheme as discussed in the SM.
The model defined by Eqs.~\eqref{eq:langevin} and \eqref{eq:collrule} has been extensively used to describe granular dynamics in different conditions \cite{Puglisi98,Sarracino2010,Plati2024}.

We define the granular temperature of the system as its average kinetic energy
\begin{equation}\label{eq:AveGranTempEDMD}
    T = \dfrac{1}{N_S + N_L}\dfrac{1}{2}\sum_{i = 1}^{N}m_i\langle \bm v_i^2\rangle,  
\end{equation}
where $\langle\dots\rangle$ is an average over multiple uncorrelated configurations in the steady state. We also define \textit{per specie} granular temperatures $T_{S}$ and $T_L$
\begin{equation}\label{eq:AveGranTempEDMDSmallBig}
    \begin{split}
    T_S &= \dfrac{1}{N_S}\dfrac{m_S}{2}\sum_{i = 1}^{N_S}\langle \bm v_i^2\rangle\\
    T_L &= \dfrac{1}{N_L}\dfrac{m_L}{2}\sum_{i = 1}^{N_L}\langle \bm v_i^2\rangle,  
    \end{split}
\end{equation}
where the sum runs only on either the small or large particles.
We note that, for $\alpha=1$, the system attains thermodynamic equilibrium at a granular temperature $T=T_b$ which is the same for the two disk species in virtue of the equipartition theorem: $T_b = T_S = T_L$. For $\alpha<1$, the dynamics is out of equilibrium but can still reach a non-equilibrium steady state with an average kinetic energy that, in principle, depends on all the control parameters of the system ${\phi,q,x_S,\alpha,\gamma,T_b}$, as well as on the phase formed by the system. Out of equilibrium, energy equipartition between the two species is no longer guaranteed \cite{Barrat2002} and the velocity probability distribution function is not Maxwellian \cite{Puglisi98}. This model contains the essential ingredients of a granular system where the temperature $T_b$ accounts for the injection of external energy and $\alpha$ for the internal dissipation due to inelastic collisions. Here we are interested in exploring how the phenomenology of the liquid-solid-like phase transition under study is affected by the non-equilibrium mechanisms regulated by $T_b$ and $\alpha$. The parameter $\gamma$, as well as masses, diameters and compositions of the mixture, are kept fixed which allows us to define an energy scale $\varepsilon=m_L(\sigma_L\gamma)^2$. Specifically, we choose $x_s=0.5$ and $q=0.476$ while the masses are fixed by requiring an equal mass density for both species.
\begin{figure}
\includegraphics[width=0.99\columnwidth,clip=true]{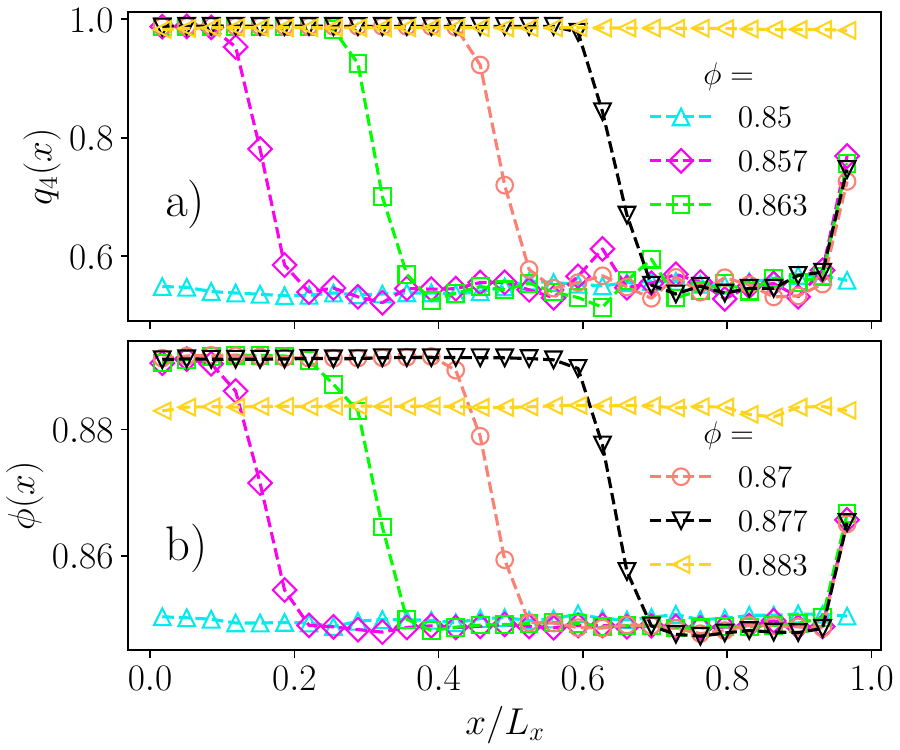}
    
    \centering
    \caption{Horizontal profiles of $q_4(x)$ and $\phi(x)$ obtained with the direct phase coexistence method implemented for the EDMD model. We report the results for six different imposed global area fractions $\phi=\{0.85 , 0.857, 0.863, 0.87 , 0.877, 0.883\}$ and $N=9600$, $\alpha = 0.94$ and $T/\varepsilon=34.175$.} 
    \label{fig:EDMD_Profile}
\end{figure}

\begin{figure*}
\centering
\includegraphics[width=0.93\textwidth,clip=true]{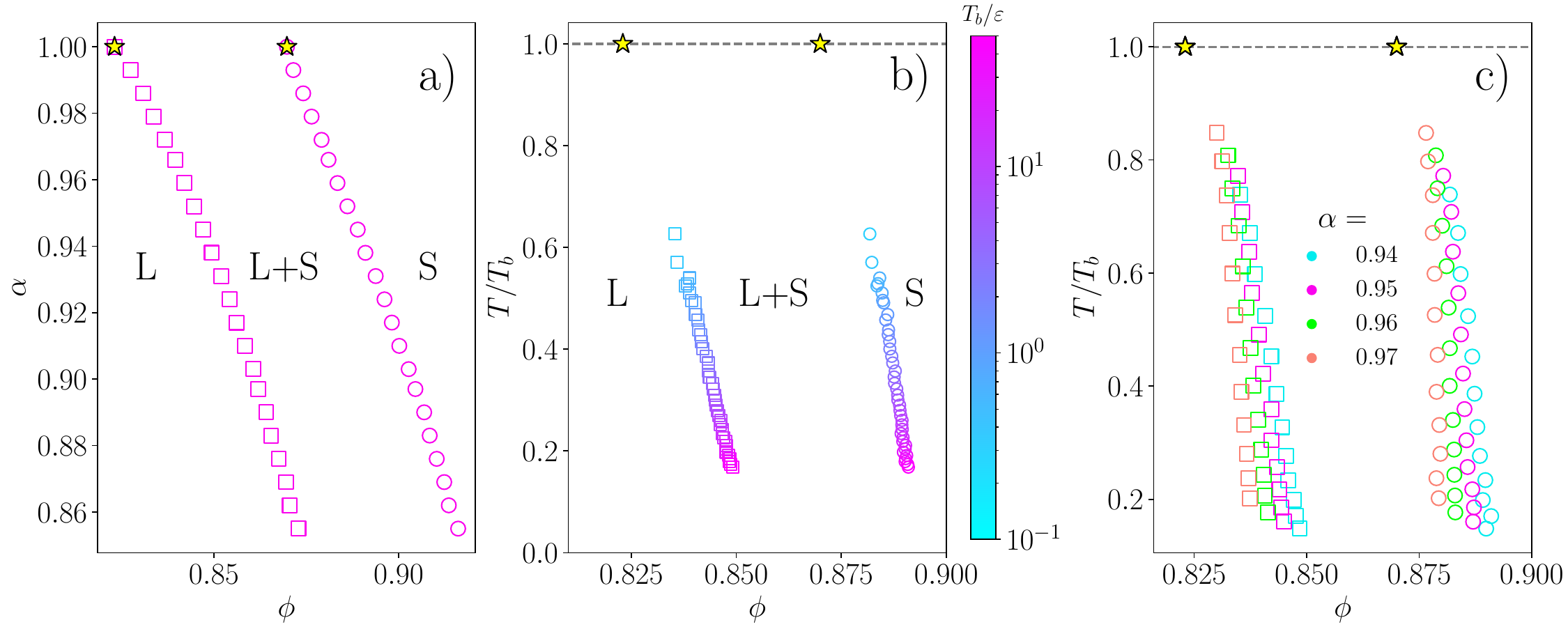}
\caption{
a) Variation of the liquid (square) and solid (circle) coexistence density as $\alpha$ is varied in the EDMD model at $T_b/\varepsilon=25$. b) Variation of the liquid (square) and solid (circle) coexistence density as $T_b$ is varied at $\alpha = 0.94$. c) Variation of the liquid (square) and solid (circle) coexistence density as $T_b$ and $\alpha$ are varied. For each plot, the stars represent the measured equilibrium coexistence density values. $N = 9600$ in all simulations.} 
\label{fig:EDMD_PhaseDiagram} 
\end{figure*}

Before exploring direct phase coexistence, we confirmed (not shown) that, for $\alpha=1$, we recover the overall phenomenology observed in Ref.~\onlinecite{Fayen2022} where the equilibrium dynamics is performed in the microcanonical ensemble (i.e. without a thermostat) and that, for $\alpha<1$, we can still observe the self-assembly of the S1 phase.  For the direct phase coexistence simulations, we used the same general setup as the one described for DEM simulations but now our simulation box is fully 2D and we slightly changed the calibration procedure to set the correct initial lattice spacing to avoid deformed crystals at coexistence (see SM). In Fig.~\ref{fig:EDMD_Profile}, we plot the $q_4(x)$ and the $\phi(x)$ profiles for a specific couple of $T_b$ and $\alpha$ observing results similar to the ones already found for DEM simulations thus validating this numerical scheme for further exploration of the granular phase coexistence. Then, we performed an extensive scan of direct phase coexistence simulations for $T_b \in [T_{min},T_{max}]$ and $\alpha \in [0.85,1]$. From each simulation, we extract $\phi_{liq}$ and $\phi_{sol}$ from the area fraction profiles so that we can build the phase diagrams reported in Fig.~\ref{fig:EDMD_PhaseDiagram}. We focus on the behavior obtained by varying $\alpha$ for fixed $T_b$ and vice-versa. 
In Fig.~\ref{fig:EDMD_PhaseDiagram}a, we can see that reducing $\alpha$ shifts the coexistence region towards larger area fractions thus promoting the stability of the granular liquid phase. Instead, in Fig.~\ref{fig:EDMD_PhaseDiagram}b, we observe that reducing $T_b$ at fixed $\alpha$ has the opposite effect promoting the stability of the S1 phase with coexistence densities getting closer to the equilibrium values.
It is interesting to note that two mechanisms that act in the same way from the energetic point of view (i.e. lowering the overall kinetic energy of the system) have a contrasting effect on the phase stability.  
In panels c) of the same figure we show the $T/T_b$ vs $\phi$ phase diagrams for all the explored values of $\alpha$ and $T_b$. 
This plot shows that the phase diagram in this representation remains qualitatively the same for different choices of $\alpha$. 
We can try to understand these observations by interpreting $T/T_b$ as a measure of the deviation from the equilibrium reference: at thermodynamic equilibrium, $T/T_b=1$.
Otherwise, the dissipative collisions ensure that $T/T_b<1$, as the energy loss of the collisions is only partly compensated for by the thermostat. At the same time, we also know that dissipative collisions can act as an effective attraction potential~\cite{Bordallo-Favela2009}. This can explain the shift of the coexistence region as a function of $\alpha$ in Fig.~\ref{fig:EDMD_PhaseDiagram}a which is consistent with what is expected in an equilibrium system, where weak and very short-ranged attractions tend to shift the coexistence densities to higher densities \cite{mederos1994phase}. 
Conversely, the results presented in Fig.~\ref{fig:EDMD_PhaseDiagram}b can be understood by considering the interplay between driving and dissipation mechanisms in the system. The Langevin bath drives the system towards equilibrium, while the dissipative collisions drive the system out of equilibrium. In the limit where $T_b \to 0$, the collision frequency between the particles approaches zero, whereas the characteristic timescale associated with the Langevin bath, $\gamma$, remains unchanged. Therefore, in this regime, the non-equilibrium effects become negligible. On the other hand, at high driving temperatures $T_b$, the system is strongly driven out of equilibrium because the frequency of the dissipative collisions scales with $\sqrt{T}$.

To summarize, the phase diagram in Fig.~\ref{fig:EDMD_PhaseDiagram}c shows how the non-equilibrium mechanisms at play in the system change its phase behavior away from that of its equilibrium hard-disk analogue.

\subsection{Temperature difference between coexisting phases}

Now that we have characterized the phase stability we focus on a remarkable non-equilibrium effect which can be observed both in the realistic and the coarse-grained model when the liquid and the S1 phase coexist. The usual picture at equilibrium is that there is no net flux of particles and heat between the two phases.
This is because any macroscopic observation of a net current is intrinsically in contradiction with the time reversibility that characterizes equilibrium dynamics. From this, it is clear that the two phases must be in equilibrium with the same temperature and the same chemical potential. In out-of-equilibrium systems, this scenario can be violated: an open system where energy constantly flows in and out (as happens in granular or active matter) can sustain the presence of net currents while its statistical properties are stationary in time. This is a key difference between non-equilibrium steady states and equilibrium dynamics. 
As a result, the coexistence of phases with different average kinetic energies has been observed in active and granular monodisperse systems \cite{Mandal2019,Hecht2024,Olafsen2005,Prevost2004,Melby2005}. At the interface, this temperature difference can give rise to a net flow of energy from the hotter to the colder phase. Here we provide evidence of this distinctive out-of-equilibrium phenomenon in direct phase coexistence simulations of our granular binary system.

\begin{figure}

   \includegraphics[width=0.99\columnwidth,clip=true]{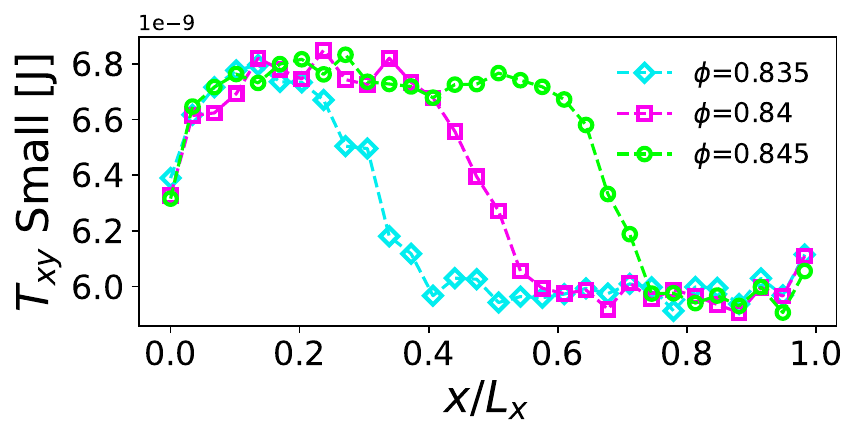}
   \includegraphics[width=0.99\columnwidth,clip=true]{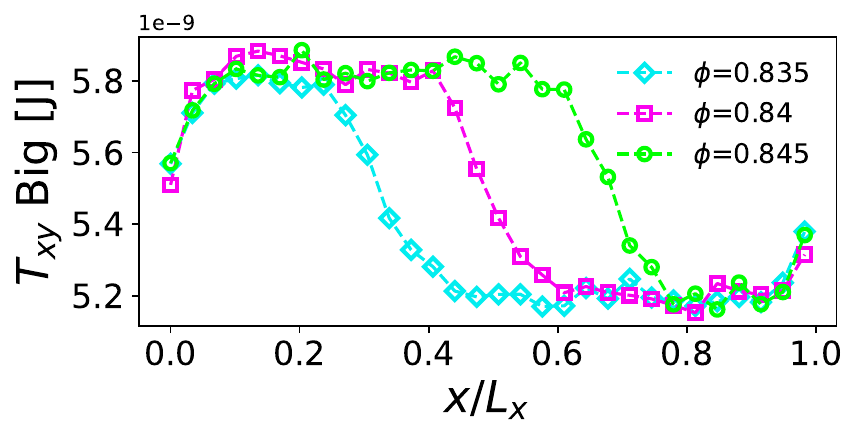}

    \centering
    \caption{Horizontal granular temperature profiles $T_{xy}(x)$ for small and large grains obtained with the direct coexistence method implemented with DEM simulations (same simulations of Fig.~\ref{fig:DEM_Profile}). 
    These simulations have been performed in a square box with PBC for $N=2000$, $q=0.476$, $x_S=0.5$, $\Gamma=4.4$ and $f = 350$ Hz.} 
    \label{fig:DEM_ProfileTEMP}
\end{figure}

\subsubsection{DEM Simulations}
In Fig.~\ref{fig:DEM_ProfileTEMP}a-b we show the horizontal profiles of the granular temperature in the $xy$-plane for small and large grains obtained through DEM simulations of the same systems analyzed in Fig.~\ref{fig:DEM_Profile}. These quantities are defined as Eqs.~\ref{eq:AveGranTempEDMDSmallBig} but considering only the horizontal projections of the grain velocities.  First of all, we note the lack of energy equipartition: small and large grains have different average granular temperatures \cite{Barrat2002}.
Comparing these profiles with the $q_4$ and area fraction ones (Fig.~\ref{fig:DEM_Profile}), it is clear that the crystal has a larger granular temperature than the fluid. We stress that this result presents a substantial difference with respect to previous studies of monodisperse systems \cite{Olafsen98,Prevost2004,Melby2005}, where the denser crystalline phase is usually ``colder" than the coexisting less dense fluid one. 
In those examples of non-equilibrium phase coexistence, the origin of the cooling in the crystalline phase was explained in terms of bistability of the grain-plate dynamics \cite{Olafsen98,Melby2005,Losert99,Geminard2003} or by the common picture for which, in granular systems, denser regions present a higher rate of dissipative collision and in turn a lower average kinetic energy \cite{Prevost2004,Melby2005}.

The scenario observed in our DEM simulations needs a different explanation and we argue that it relies on the interplay between $z$-to-$xy$ energy transfer and the local structure of the system. Indeed, apart from some specific mechanisms \footnote{These are: impacts with surface asperities (experiments) and coupling between rotational and translational dynamics mediated by the tangential friction with the plate (experiments and DEM simulations).}, it is reasonable to think that the main energy transfer from the plate to the granular system occurs in the vertical direction. Then, collisions between grains at different heights allow for a subsequent energy transfer from $z$ to the horizontal plane.
Such off-planar collisions are particularly favoured between spheres of different sizes because their centers are on average at different heights when they vibrate on a substrate. In turn, the S1 structure favours contact between small and large grains, so the $z$-to-$xy$ energy transfer is expected to be more efficient in this phase than in the liquid. 
We believe that this is the most relevant effect explaining the coexistence between a ``hot" crystal and a ``cold" fluid in our quasi-2D granular system.

\subsubsection{EDMD Simulations and Kinetic Theory}
\begin{figure}
   \includegraphics[width=0.99\columnwidth,clip=true]{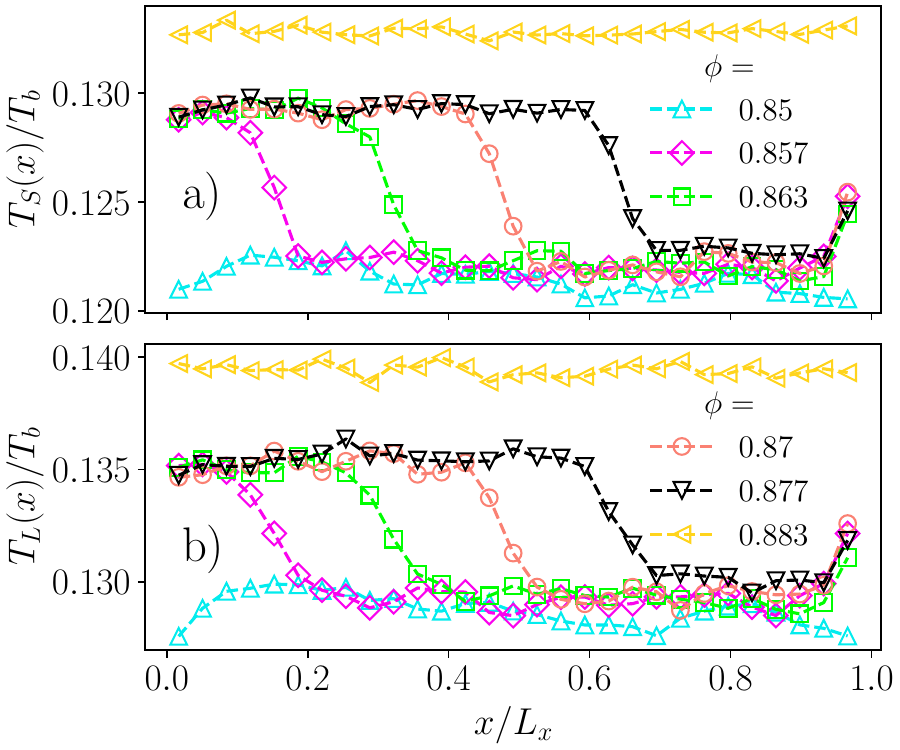}
    \centering
    \caption{Horizontal profiles of $T_S$ and $T_L$ obtained with the direct phase coexistence method implemented with EDMD simulations. We report the results for six different imposed global area fractions $\phi \in\{0.85 , 0.857, 0.863, 0.87 , 0.877, 0.883\}$ , $N=9600$, $\alpha = 0.94$ and $T/\varepsilon=34.175$.} 
    \label{fig:EDMD_ProfileTEMP18}
\end{figure}

Coming back to the EDMD simulations, in Fig.~\ref{fig:EDMD_ProfileTEMP18} we plot the granular temperature profiles of small and big particles for the simulations analyzed in Fig.~\ref{fig:EDMD_Profile}.  The same phenomenology discussed for DEM simulations is recovered, namely a hot S1 crystal coexisting with a cold liquid.
Additionally, here we note that the undercompressed crystal at $\phi=0.883$ (i.e. at a area fraction below the melting point) is hotter than the coexisting crystals. This observation aligns with the usual picture for which, within a single phase, the granular temperature is mainly governed by the overall density~\cite{Prevost2004,Melby2005}.

Despite the similar phenomenology obtained in EDMD and DEM, we cannot use the same argument about optimization of the energy transfer given by off-planar particle collisions since in the EDMD simulations, energy is directly injected in the $xy$ plane by the thermostat and collisions are dissipative events. Based on these premises, and knowing that the crystal is denser than the fluid, one would expect to find a crystalline phase colder than the coexisting fluid which is clearly in contradiction with our numerical results. To understand the behavior of the granular temperature observed in EDMD simulations we adapted the kinetic theory for granular mixtures developed by Barrat and Trizac \cite{Barrat2002} to our collisional model for non-additive disks defined by Eqs.~\eqref{eq:langevin} and~\eqref{eq:collrule} obtaining the following equations for the granular temperatures $T_i$ of each species $i\in\{S,L\}$:
\begin{widetext}
\begin{equation}\label{eq: kinTheoBinary}
  \frac{ \sqrt{\pi}\gamma(T_b-T_i)}{2m_i}=\chi_{ii}\phi_i \frac{(1-\alpha^2)}{\sigma_i m_i^{3/2}}T_i^{3/2}+\chi_{ij}\phi_j \sqrt{\frac{\sigma_i}{\sigma_j^3}}\mu_{ji}^2\left[(1-\alpha^2)\left( \frac{2T_i}{m_i}+\frac{2T_j}{m_j}  \right) + 4(1+\alpha)\frac{T_i-T_j}{m_j}\right]\sqrt{\left( \frac{2T_i}{m_i}+\frac{2T_j}{m_j}  \right)},
\end{equation}
\end{widetext}
where $\phi_i=\pi\sigma_i^2/(4L_xL_y)$, and $\chi_{ij}$ are the pair distribution functions at contact between species $i$ and $j$. In this notation, when $i=S$, then $j=L$, and conversely. We point out that the latter quantities are not known a priori and generally depend on $T_i$, $T_j$, $\phi_i$ and $\phi_j$. We do not enter here into the details of the derivation, for which we refer to \cite{Barrat2002}, but we only point out that the left-hand-side of the above equation accounts for the effect of the thermal bath on species $i$ while the right-hand-side expresses the average dissipation rate. Equating these two contributions gives the condition that must be fulfilled by the granular temperatures $T_S$ and $T_L$ in the non-equilibrium steady state.
Of course, Eq.~\eqref{eq: kinTheoBinary} does not directly provide us with an explicit analytical expression of $T_S$ and $T_L$ as a function of the control parameters. However, it tells us that the quantities that govern the granular temperatures of the two species in a binary mixture are the four products given by the terms $\chi_{ii}\phi_i$ and $\chi_{ij}\phi_j$. This already highlights the non-trivial interplay between local structure and global density in the energetic balance between forcing and dissipation. 
It is important to stress that Eq.~\eqref{eq: kinTheoBinary} has been obtained following the standard scheme of kinetic theory~\cite{BrilliantovBook,Barrat2002}, then assuming molecular chaos and a Maxwellian velocity distribution for the grains. Both these assumptions are never exactly fulfilled in a granular material but, in dilute systems and for $\alpha$ close to 1, the predictions of kinetic theory are usually in good agreement with simulation results~\cite{Barrat2002}. Our simulations are performed at $\alpha=0.94$ but, to probe the 
liquid-solid-like transition, they explore a range of very high densities. 
\begin{figure*}
   \includegraphics[width=0.93\textwidth,clip=true]{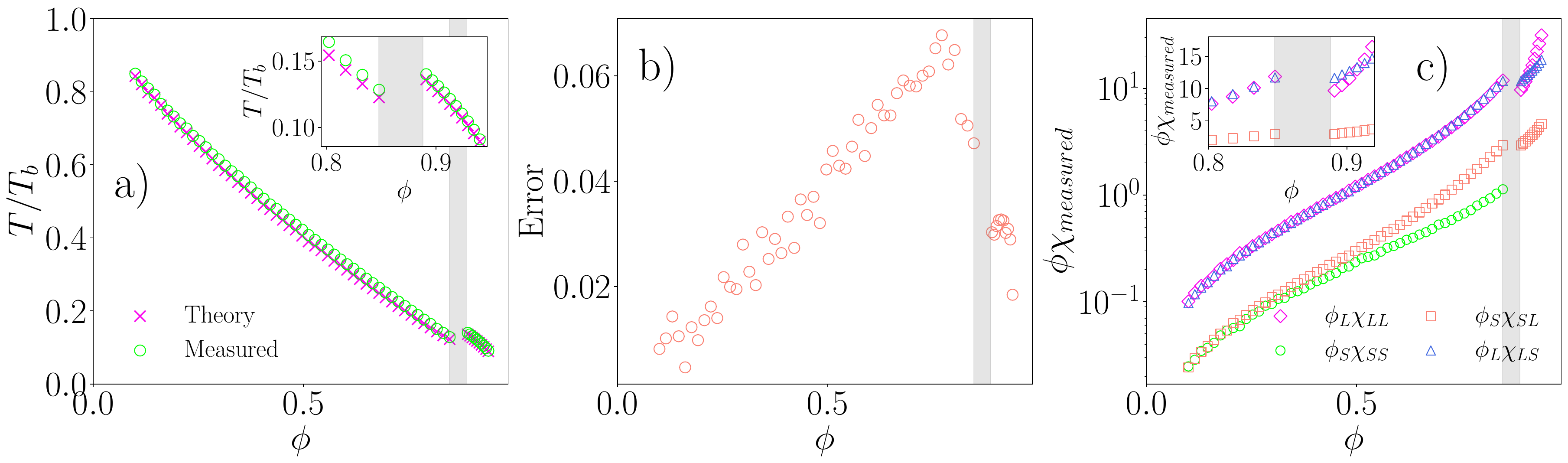}
    \centering
    \caption{Kinetic theory for the total temperature in a binary mixture in the liquid and in the S1 crystal. a)  Measured values of $\phi_i \chi_{ij}$ with respect to $\phi$in the liquid and the solid.  Inset is a zoom around the coexistence region. Note that in the solid $\chi_{ss}=0$ because in a defect-free $S1$ configuration, pairs of small particles never touch. b) Comparison as a function of the area fraction, of the measured and predicted total temperature of the system (Eq.~\ref{eq: kinTheoBinary}). The grey area marks the coexistence region between the liquid and the solid. Inset is a zoom around the coexistence region.  c) Error between the measured temperature and the theoretical prediction $(T_\mathrm{measured}-T_\mathrm{theory})/T_\mathrm{theory}$ as a function of the area fraction.  Similar results are obtained for $T_{S}$ and $T_{L}$ alone. $N = 2\times 223^2$, $\alpha = 0.94$ and $T/\varepsilon=34.175$.} 
    \label{fig:EDMD_theory_temp}
\end{figure*}
To test the applicability of these theoretical predictions to our simulations, we compared the value of the average granular temperature $T$  (Eq.~\eqref{eq:AveGranTempEDMD}) obtained with kinetic theory (i.e. solving numerically the system of two equations given by Eq.~\eqref{eq: kinTheoBinary}) with the one of EDMD simulations for a wide range of area fractions spanning from the dilute gas regime to the solid state. We point out that to find the solution from kinetic theory we must insert in Eq.~\eqref{eq: kinTheoBinary} the values of $\chi_{ij}$ measured from simulations. 
The results of this analysis are reported in Fig.~\ref{fig:EDMD_theory_temp}a and b where we first note a good agreement between kinetic theory and simulations for the whole range of explored area fractions.  Then, we also observe that the granular temperature is always a decreasing function of the area fraction except around the liquid-solid phase transition. Here we see that there exists a range of area fractions where the crystal is hotter than the liquid despite being denser. Even if the kinetic theory cannot be used to directly predict the behavior of the system inside the coexistence region, we can take the values of the granular temperatures at the right and left edge of it as representative of respectively the liquid and the crystalline ones at coexistence. This means that kinetic theory is able to predict that the solid must be hotter than the liquid. 
An explanation for this behavior in the 2D model can be found by looking at the products $\chi_{ii}\phi_i$ and $\chi_{ij}\phi_j$ plotted in Fig~\ref{fig:EDMD_theory_temp}c. Here we see that they are monotonously increasing with $\phi$ apart from around the coexistence region where the large-large and the large-small contributions experience a jump down and  $\phi\chi_{SS}$ becomes zero (as we expect in the S1 structure). From the way these products enter the right-hand-side of Eq.~\eqref{eq: kinTheoBinary} (which represents the dissipation rate), we can understand that the observed behavior of $\chi_{ii}\phi_i$ and $\chi_{ij}\phi_j$ suggests that the transition from the solid to the S1 structure can reduce the average dissipation in the system. Indeed, the dissipative contributions of $LL$- and $LS$-collisions are reduced while collisions between small grains are simply not possible in the S1 crystal.

The performed analysis suggests that within a single phase, the common scenario for which higher densities lead to an increase in the dissipation rate and in turn to a lower kinetic energy is applicable. However, when the fluid-crystal-like transition occurs, the effect of the local structure sets in and the granular temperature is primarily determined by the products between the global area fraction and the contact values of the pair distribution function. This allows the possibility of having denser granular crystals with a higher kinetic energy than less dense granular liquids.

Finally, to avoid possible confusion, we stress again here that our interpretation of phase coexistence at different granular temperatures in DEM and EDMD is based on two independent mechanisms that rely on the details of the system under study (e.g. off-planar collisions in DEM). There are many ways to introduce the effect of forcing, dissipation and confinement in granular models and we expect the temperature in the solid to be larger or smaller than the one in the liquid depending on these choices. Nevertheless, we believe that the physical insights provided in this section offer two concrete examples of how the coupling between local structure and energy flows can originate a stable kinetic temperature difference between coexisting phases out of equilibrium.

 \section{Conclusions} \label{sec: conc}

In this study, we have shown experimentally that a granular binary mixture of spherical grains vibrated on a substrate undergoes a fluid-crystal-like phase transition between a disordered phase and a square binary crystal. The transition behaves remarkably similar to an equilibrium fluid-crystal transition, including the observation of a stable coexistence between the two phases. To further investigate this experimental observation, we used computer simulations of both a realistic granular model and a coarse-grained collision-based model. In both models, we again observe strongly equilibrium-like phase coexistences, obeying the lever rule.  However, the similarity to equilibrium phenomenology is broken when looking at the granular temperature, which is different in the two coexisting phases. Using a kinetic theory for binary mixtures, we show how this temperature difference arises from the interplay between local structure and energy transfer mechanisms. 
\\
Our study demonstrates how crystal self-assembly can be controlled based on equilibrium predictions even in the presence of strong non-equilibrium effects. In particular, the case of binary granular crystals offers a way to realize mixtures with a spatially constant composition. Additionally, our results highlight how simple kinetic theories can shed light on the physics of non-equilibrium phase coexistences.

\begin{acknowledgments}

We thank B. Darbois Texier for helping us find the granular materials for the experiment and for the precious feedback on our early results, we also thank A. Puglisi, A. Gnoli and E. Fayen for their invaluable support in setting up this project and S. Cabaret for the design of the quasi-2D cell. This work has been carried on with the support of Investissements d'Avenir of LabEx PALM (Grant No. ANR-10-LABX-0039-PALM) and funding
from the Agence Nationale de la Recherche (ANR), grant ANR-21-CE06-0039. Two of the authors (RM and GF) partially wrote this paper while participating in a program at the Erwin Schrödinger Institute (ESI) in Vienna.

\end{acknowledgments}

\bibliographystyle{apsrev4-1}
\bibliography{Biblio2}

\end{document}